\documentclass{aastex631}
\usepackage{multirow}

\newcommand{\hii}{H~{\small II~}}

\shorttitle{Spectral-line survey of W51e1/e2}
\shortauthors{Kalenskii et al.}
\begin{document}

\title{Spectral-line Survey of the Region of Massive Star Formation W51e1/e2 in the 4 mm Wavelength Range}

\author[0000-0002-4294-3643]{Sergei. V. Kalenskii}
\affiliation{Lebedev Physical Institute, Astro Space Center,\\ 84/32 Profsoyuznaya st., Moscow, GSP-7, 117997, Russia}

\author{Ralf I. Kaiser}
\affiliation{Department of Chemistry, University of Hawaii at Manoa, Honolulu, HI, 96822, USA}

\author{Per Bergman}
\affiliation{Department of Space, Earth and Environment, Chalmers University of Technology,\\ Onsala Space Observatory, SE-43992 Onsala, Sweden}

\author{A. O. Henrik Olofsson}
\affiliation{Department of Space, Earth and Environment, Chalmers University of Technology,\\ Onsala Space Observatory, SE-43992 Onsala, Sweden}

\author{Kirill D. Degtyarev}
\affiliation{Moscow Institute of Physics and Technology, 9 Institutskiy per., Dolgoprudny, Moscow Region, 141700, Russia}
 
\author{Polina Golysheva}
\affiliation{Lebedev Physical Institute, Astro Space Center,\\ 84/32 Profsoyuznaya st., Moscow, GSP-7, 117997, Russia}
\affiliation{Sternberg Astronomical Institute, Moscow State University, \\ Universitetsky pr., 13, Moscow 119234, Russia}

\begin{abstract}

We present the results of a spectral-line survey of the W51e1/e2 star-forming region at 68 -- 88 GHz. 79 molecules and their isotopologues were detected, from simple diatomic or triatomic molecules, such as SO, SiO, and CCH, to complex organic compounds, such as CH$_3$OCH$_3$ or CH$_3$COCH$_3$. A number of lines that are absent from the Lovas list of molecular lines observed in space were detected, and most of these were identified. A significant number of the detected molecules are typical for hot cores. These include the neutral molecules HCOOCH$_3$, CH$_3$CH$_2$OH, CH$_3$COCH$_3$ etc, which are currently believed to exist in the gas phase only in hot cores and shock-heated gas. In addition, vibrationally excited C$_4$H and HC$_3$N lines with upper-level energies of several hundred Kelvins were found. Such lines can arise only in hot gas with temperatures on the order of 100 K or higher. Apart from neutral molecules, various molecular ions were also detected. Some of these (HC$^{18}$O$^+$, H$^{13}$CO$^+$, and HCS$^+$) usually exist in molecular clouds with high visual extinctions. Potential formation pathways of complex organic molecules (COMs) and of hydrocarbons, along with nitriles, are considered. These formation routes are first discussed in the context of laboratory experiments elucidating the synthesis of organic molecules in interstellar ices in cold molecular clouds, followed by sublimation into the gas phase in the hot core stage. Thereafter, we discuss the predominant formation of hydrocarbons and their nitriles in the gas phase through bimolecular neutral--neutral reactions.

\end{abstract}


\section{Introduction} \label{sec:intro}
Spectral-line surveys covering wide frequency bands are an effective tool for studying molecular clouds. Since multiple spectral lines fall within the ranges of such surveys, they allow us to fully investigate the molecular compositions of the studied sources and to explore the chemical processes taking place there. Surveys performed with high sensitivity make it possible to search for molecules previously undetected in space. For example, as a result of the survey of the dark cloud TMC-1, performed by~\citet{2004PASJ...56...69K}, 11 new molecules were discovered; in addition, using the spectra obtained in this survey, it was later possible to detect a 13-atom molecule, c-C$_6$H$_5$CN~\citep[benzonitrile;][]{2017arXiv170806829K,2018Sci...359..202M}. In 2018, a new, more sensitive spectral survey of TMC-1 was launched, which was named GOTHAM~\citep{2020ApJ...900L..10M}. As a result of this still-incomplete survey, about a dozen new molecules have been discovered, including  two nitrile-group-functionalized polycyclic aromatic hydrocarbons, 1- and 2-cyanonaphthalene \citep[C$_{10}$H$_7$CN;][]{2021Sci...371.1265M}.

In addition to establishing the molecular compositions of cosmic sources, spectral surveys make it possible to explore their structures and physical parameters. For example, according to the results of the scanning the region of high-mass star formation closest to the Sun, OMC-1,~three subregions were identified, designated Ridge, Plateau, and Hot Core, and  the basic physical parameters and molecular compositions of each of them were determined~\citep{1987ApJ...315..621B}.

In 2010, the results of the spectral scanning of the regions of high-mass star formation DR21OH and W51e1/e2, in the 3 mm wavelength range, performed with the 20 m radio telescope of the Onsala Space Observatory (Sweden), were published~\citep{2010ARep...54..295K, 2010ARep...54.1084K}. In W51e1/e2, the emission of 105 molecules was detected, from simple diatomic or triatomic species, such as CO, CS, and HCN, up to complex organic compounds, such as CH$_3$OCH$_3$,  CH$_3$COCH$_3$, or HCOOC$_2$H$_5$. A significant number of the detected molecules are typical for hot cores, for example, HCOOCH$_3$, CH$_3$CH$_2$OH, and CH$_3$COCH$_3$. In addition, lines of the vibrationally excited states of SiO, C$_4$H, HCN, l-C$_3$H, HC$_3$N, CH$_3$CN, CH$_3$OH, H$_2$O, and SO$_2$ were detected, with upper-level excitation temperatures on the order of several hundred Kelvins. Such lines can occur only in hot gas with a temperature on the order of 100~K or higher. An interesting result was the possible detection of two molecules, MgCN and NaCN, which have so far been observed only in the atmospheres of evolved giant stars \citep{1995ApJ...445L..47Z,2000AAS...196.0511H}. Further observations are needed to confirm or refute this detection. In addition, the stacked spectra suggest the existence of several molecules not yet discovered in space --- HOCH$_2$COOH (glycolic acid), H$_2$SO$_4$ (sulfuric acid), gG'a-CH$_3$CHOHCH$_2$OH (propanediol), and l-C$_7$H$_2$ (heptahexainylidine)\footnote{The possible detection of these molecules was not mentioned in~\citet{2010ARep...54.1084K}}.

Therefore, when a new 4 mm wavelength receiver was put into operation at Onsala, we conducted spectral scanning of DR21OH and W51e1/e2 in the 4 mm wavelength range as well. The main purpose of the new survey was to observe molecules whose lines did not fall into the band of the previous survey (DCO$^+$, DCN, DNC, etc), and to check possible detections of molecules made as a result of the previous survey. This article is devoted to W51e1/e2; the results for DR~21~OH will be published separately.

\subsection{Region of massive star formation W51e1/e2.}
\label{subsec:w51}
\begin{figure}
    \includegraphics[width=0.5\textwidth]{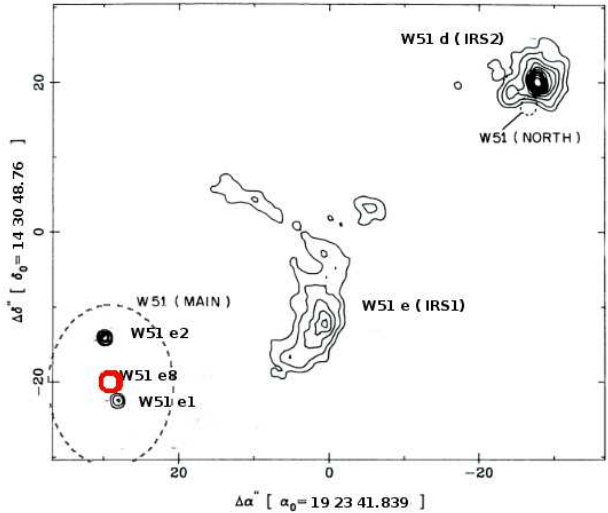}
    \caption{Solid contours: 15~GHz continuum emission in the W51 region, taken from~\citet{1987ApJS...65..193G}. The red circle shows the hot core e8. The dashed ellipse denotes the region W51~Main.}
    \label{fig:w51map}
\end{figure}

The complex of H {\small II} regions W51 is located in a giant molecular cloud in the Sagittarius arm at a distance of $5.41^{+0.31}_{-0.28}$ kpc, very close to the tangent point of the Sagittarius spiral arm~\citep{2010ApJ...720.1055S}, and it consists of four components --- A,B,C, and D~\citep{kundu}. W51A consists of eight components in the centimeter continuum, designated W51 a--h. The strongest components are e and d~\citep{martin}. \citet{scott} identified two closely located sources radio continuum sources in the vicinity of the component e, designated e1 and e2. The whole region around these two sources is called W51e1/e2. In addition to thermal radio emission, the evidence for an actively ongoing process of star formation in W51A is the presence of bright IR sources, powerful OH, H$_2$O, and CH$_3$OH masers, as well as of hot cores with rich molecular compositions. The bolometric luminosity of this complex is about $2\times 10^7$~$L_\odot$, which corresponds to at least 20 O8 stars~\citep{gins16}. These stars are associated with three sources: W51~Main, W51~IRS1, and W51~IRS2 (also called W51 North; see Fig.~\ref{fig:w51map}). The IR sources and centers of maser activity W51~Main and W51~IRS2 are associated with the components e1/e2 and d, respectively. These objects are the regions of high-mass star formation in the early stages of evolution. W51 IRS1 is associated with a much more evolved H~{\small II} region. 

Very Large Array observations have shown that W51 Main is a cluster of UC and HC \hii regions designated e1--e8~\citep{gaume93,1994ApJS...91..713M,zhang97,2016A&A...589A..44G}. The total bolometric luminosity of the cluster is on the order of $10^7$~$L_\odot$~\citep[see, e.g.][and references therein]{2016A&A...589A..44G}. The brightest HC \hii region is e2, which is ionized by an O8 star. It consists of three young stellar objects: W51e2-W, which manifests itself primarily as a hypercompact \hii region, W51e2-E, which is the brightest source of dust radiation, as well as the center of a bipolar outflow; and W51e2-NW, which is the center of bipolar outflow detected by the proper motion of H$_2$O masers. The mass of e2 is 96~$M_\odot$ \citep{2014ApJ...786...38H}. The most prominent hot cores in W51e1/e2 are associated with e2 and the center of maser activity e8, which is located at a distance of $\sim 6"$ south of e2 and contains $\sim 70$~$M_\odot$ of hot 
molecular gas. In these two regions, a large number of complex molecules characteristic of hot cores, e.g., HCOOCH$_3$, CH$_3$COOH, CH$_3$CH$_2$OH etc have been detected. LSR radial velocities of the molecular lines in e2 are about 55--57~km/s, and in e8, about 58--60~km/s. The e2 region hosts OH, H$_2$O, CH$_3$OH, and CS masers, and the e8 region OH and H$_2$O masers~\citep[][and references therein]{2019AJ....158..208G}.  \citet{2016A&A...589A..44G} mapped e2 and e8 in metastable lines of ammonia (6,6), (7,7), (9,9), (10,10), and (13,13), with  upper-level temperatures $E_u/k$ within the range 408--1691~K. Their maps demonstrate the decrease of the core image diameters with the increase of the upper-level energy. On the other hand,~\citet{1998ApJ...494..636Z} found a  kinetic temperature as low as $\sim 40$~K in the outer regions of e2 and e8. Thus, the kinetic temperature of e2 and e8 varies widely, decreasing with the increase in distance from the protostars.

\section{Observations} \label{sec:obs}
The observations were carried out with the 20 m radome-enclosed millimeter-wave radio telescope of the Onsala Space Observatory (Sweden) at 68--88~GHz. 20 observation cycles were performed in 2018 April--May. The receiver frontend was a dual-polarization sideband-separating HEMT mixer. A dual-beam switching mode with a switching frequency of 2~Hz and a beam throw of $11'$ in azimuth was applied. The half-power beamwidth (HPBW) at 86 GHz was $43"$. The antenna was pointed toward e2 (R.A.=$19^h23^m43.89^s$, decl.=$14^{o}30'36.4''$; J2000). In this case, the whole W51e1/e2 region falls within the antenna beam. Pointing errors were checked using observations of SiO masers at 86~GHz, and found to be within $5"$. The main-beam efficiency varied slightly with frequency and elevation, and was about 0.57 at 86~GHz and an elevation around $65^o$. The data were calibrated using the chopper-wheel method. The system noise temperature mostly varied within the range $\sim 200 - 500$~K, but in the last four observing cycles it varied from $\sim 300$~K to $\sim 1200$~K.

A serious problem was the fast radome-induced ripple in the spectral domain, which is more severe in the 4 mm band than the 3 mm band due to the radome material frequency response. This ripple was reduced by using the Path Length Modulator (PLM), which moves the receiver exactly one wavelength along the telescope axis during each integration.

The backend was a fast Fourier transform spectrometer, consisting of four sections. Two sections were used for each plane of polarization, providing an analysis band of 4~GHz. The frequency resolution was 76.294~kHz, corresponding to a velocity resolution of 0.28~km~s$^{-1}$ at 83~GHz. At this resolution, the achieved $1\sigma$ noise level in the antenna temperature was about 0.012~K at the lower end of the observed frequency range, smoothly decreased while the frequency increased and reached a value of $\sim 0.006$~K at 70 GHz, then remained almost constant up to a frequency of 85 GHz, before further increasing with frequency, reaching $\sim 0.008$~K at the upper boundary of the observed range. During the data reduction, the frequency resolution was smoothed to $0.305$~MHz (1.1~km~s$^{-1}$ at 83~GHz), so that the rms decreased to 0.003~K for 75\% of the band of the survey. 

This project served as a pathfinder for the new Bifrost control system, which can conduct observations in a highly automated fashion. Scripts were written that chose science sources with regard to visibility and strived to meet preset observation time targets that varied with tuning, due to the changing atmospheric transmission. Interruptions for pointing and focusing, which also involved automatic retunings, were done based on the time triggers and flag variables, and out of a small number of preselected (at the time) sufficiently strong pointing sources, one was selected by the software based on its current sky position. Observations were halted automatically when weather-related observing conditions increased the system noise above a certain limit (and resumed once the noise decreased again).

Data were reduced using the CLASS software, which is a part of the GILDAS package of the Grenoble observatory.
\begin{center}
\begin{figure}
\includegraphics[width=0.45\textwidth]{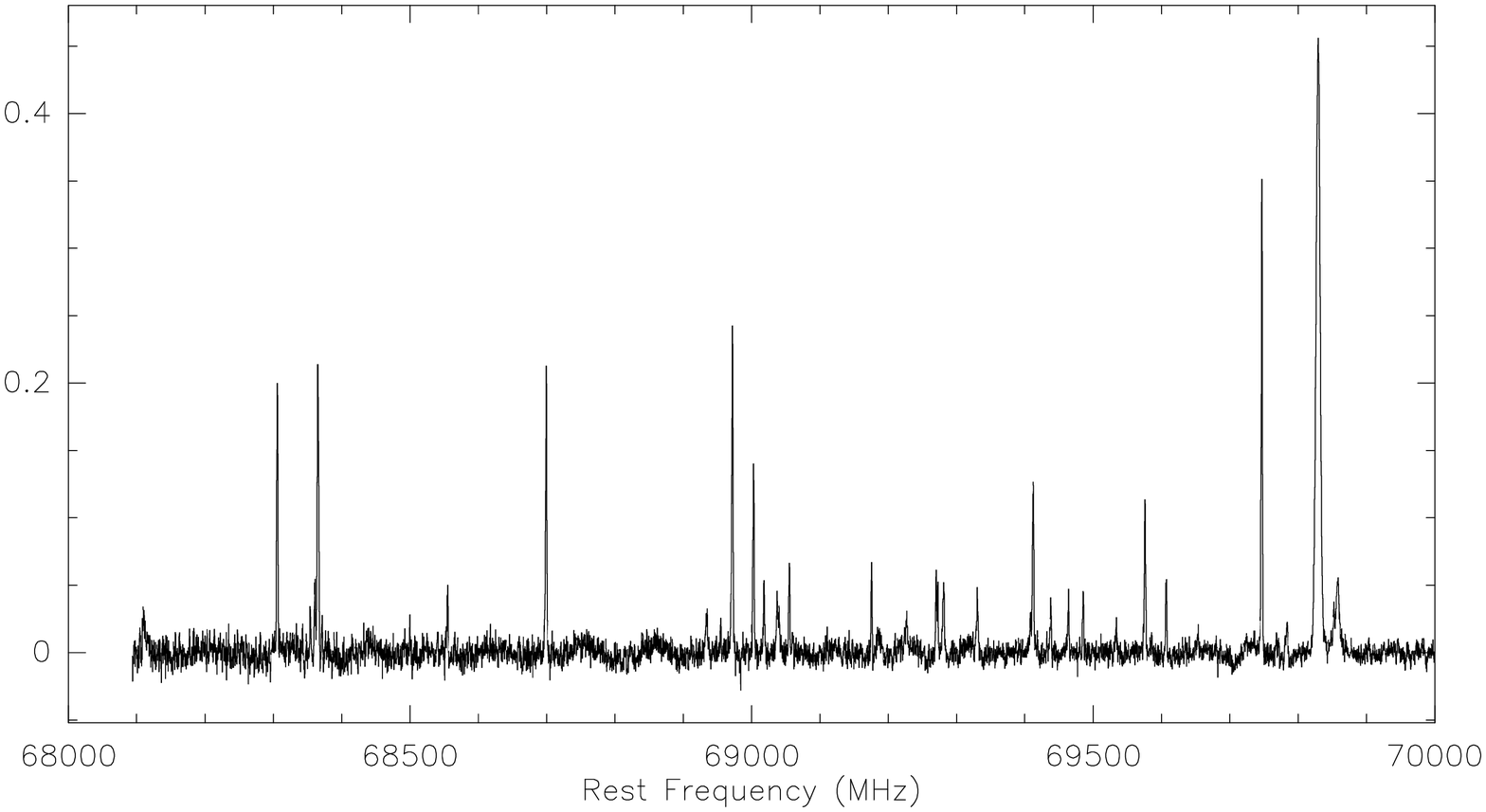}
\includegraphics[width=0.45\textwidth]{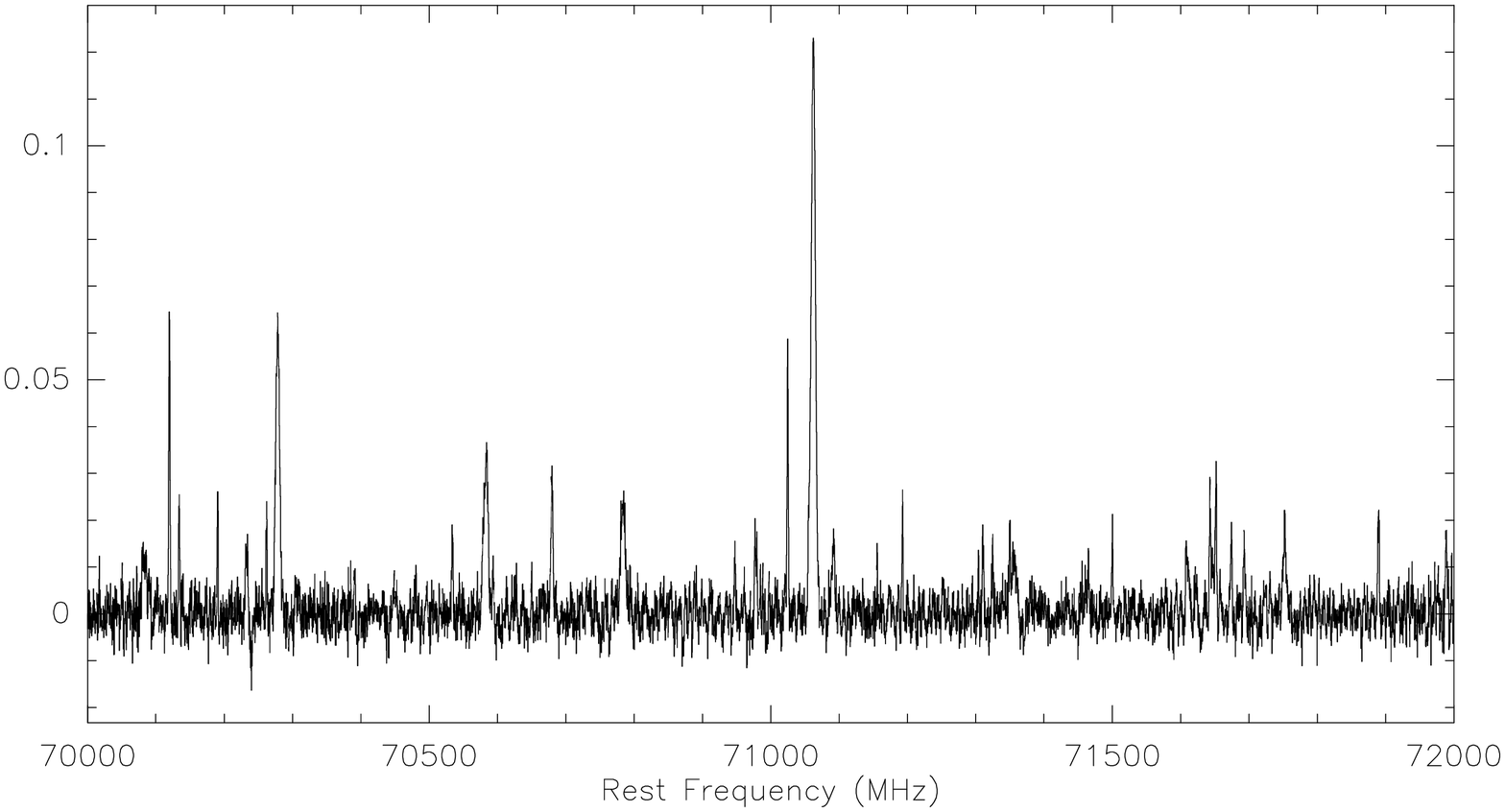}

\includegraphics[width=0.45\textwidth]{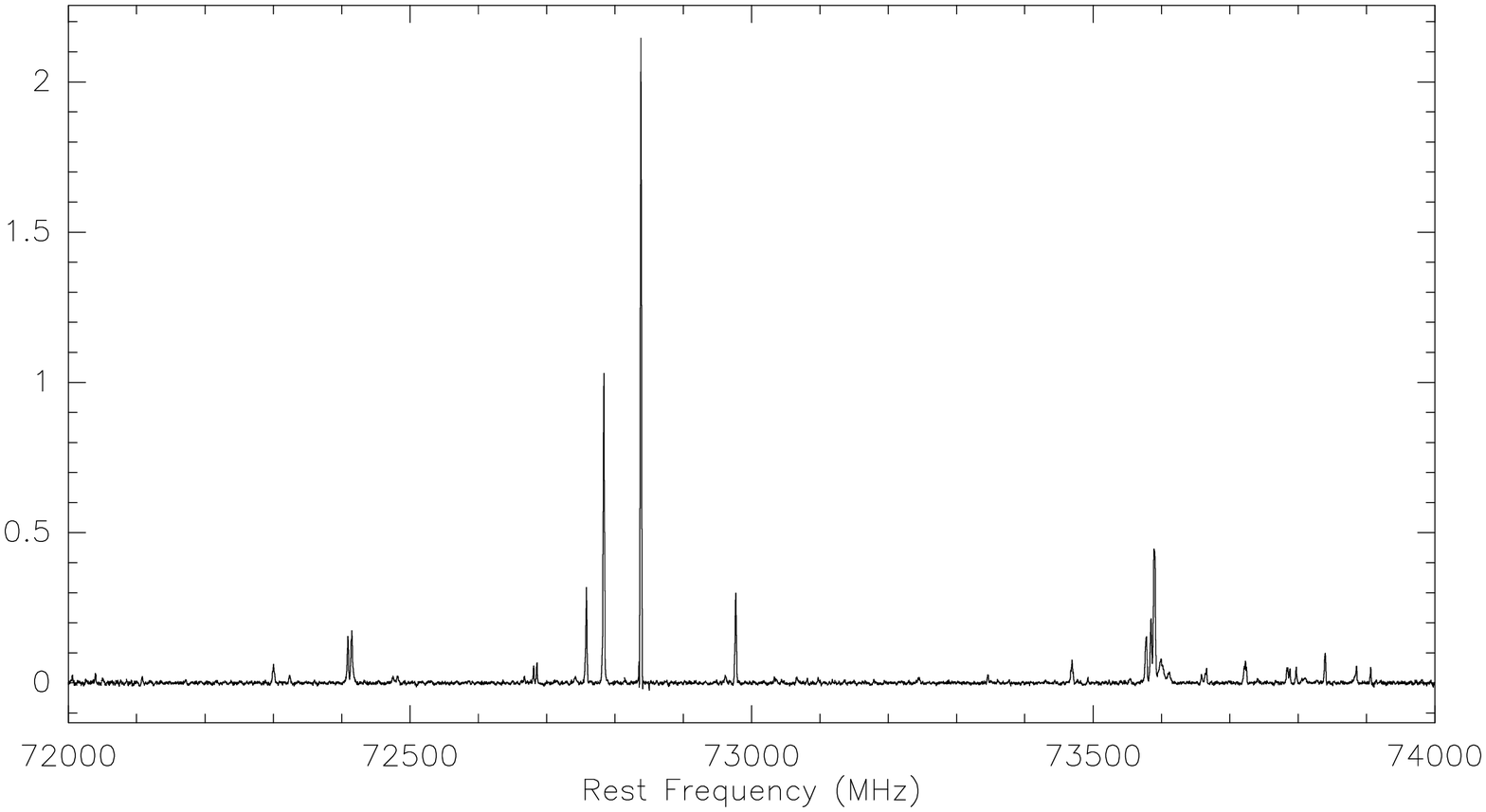}
\includegraphics[width=0.45\textwidth]{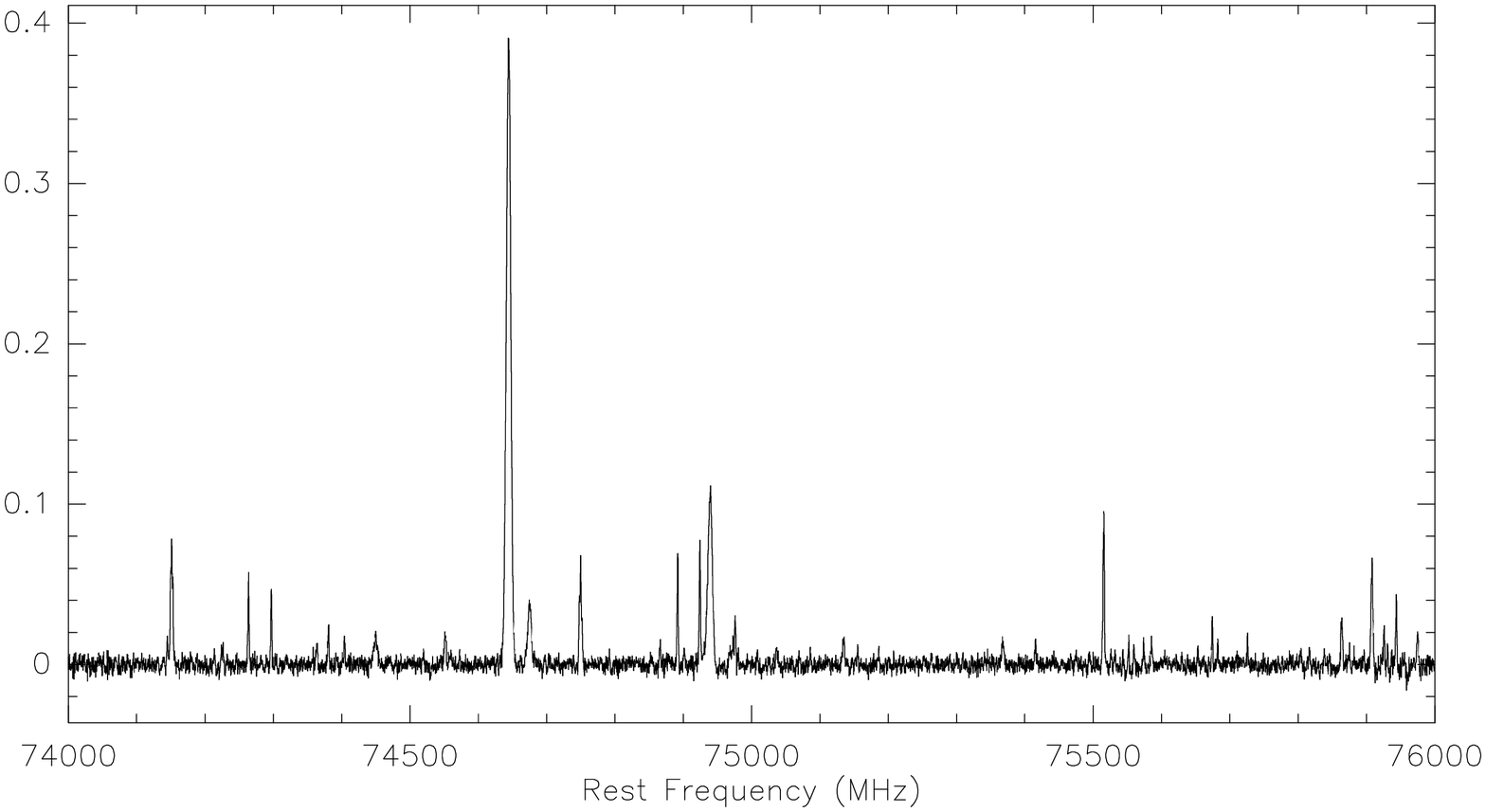}

\includegraphics[width=0.45\textwidth]{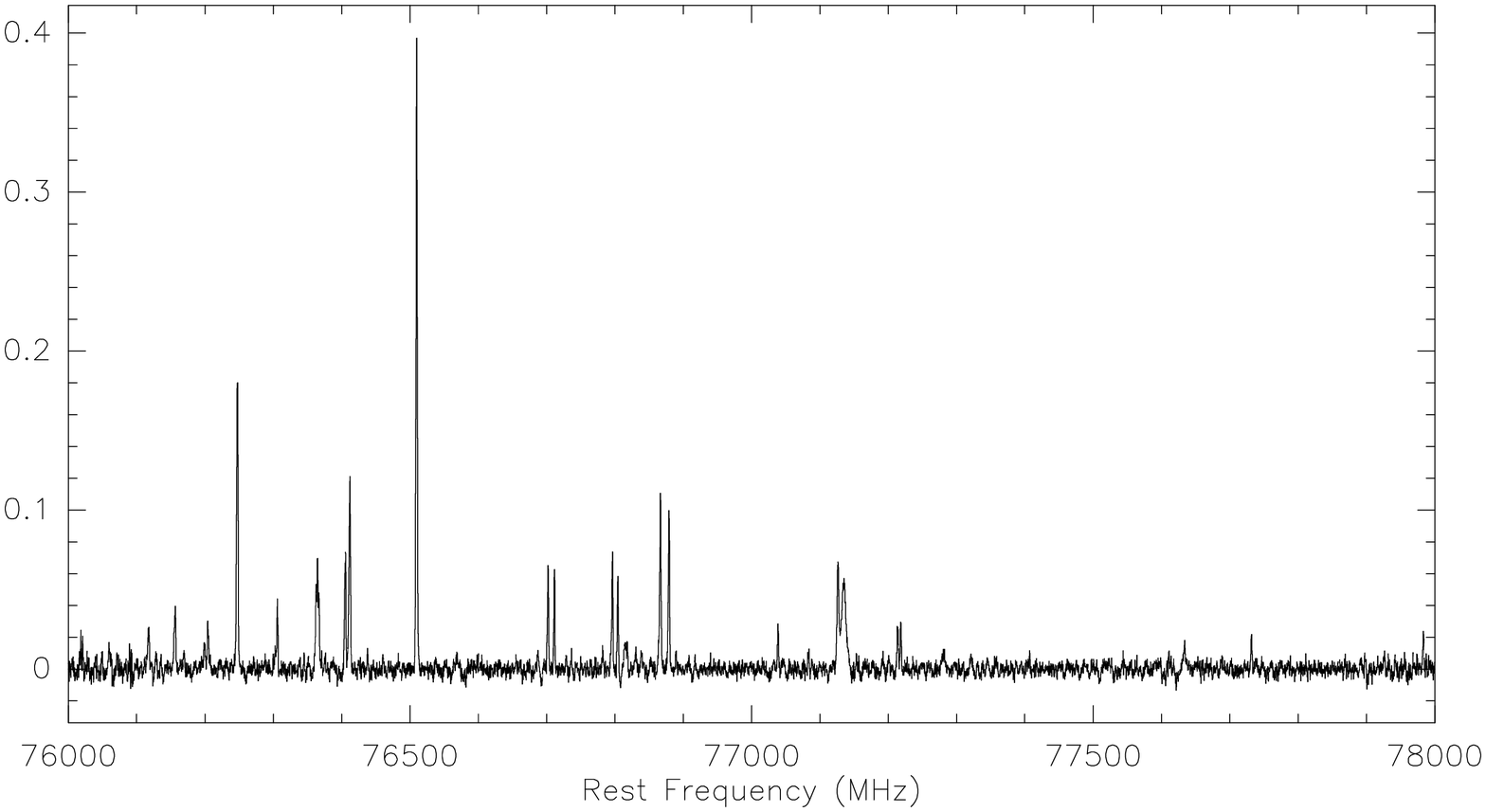}
\includegraphics[width=0.45\textwidth]{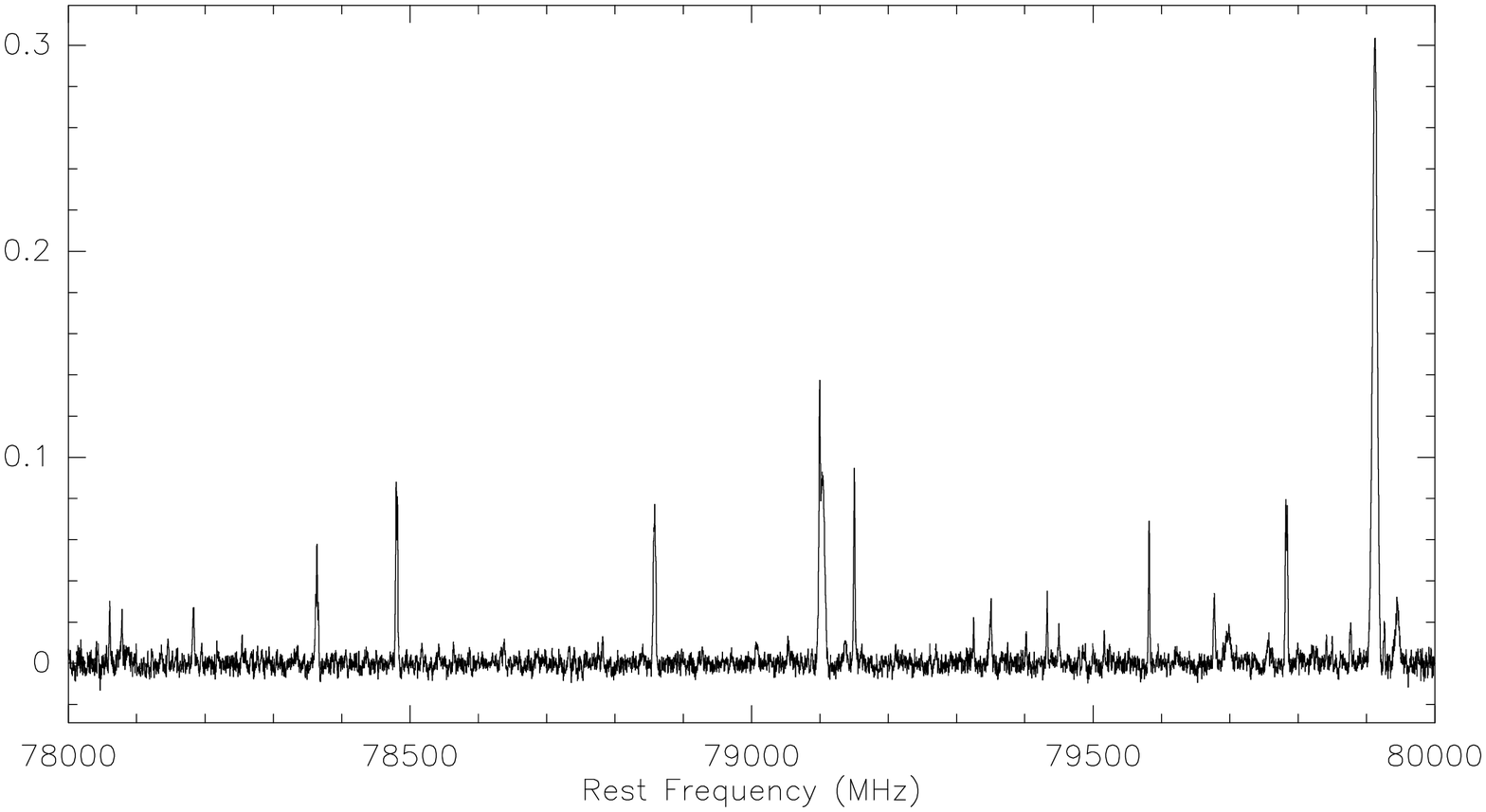}

\includegraphics[width=0.45\textwidth]{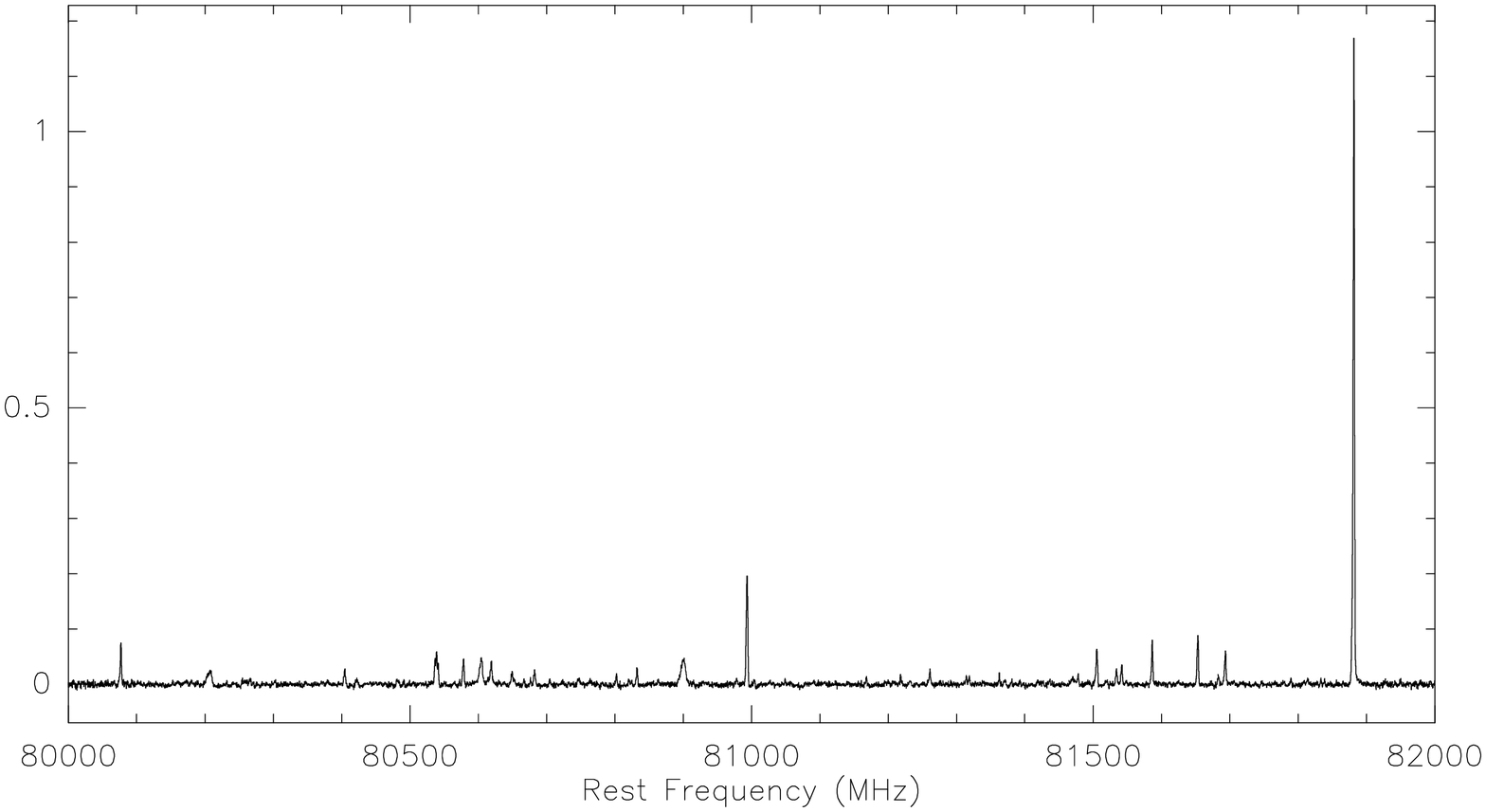}
\includegraphics[width=0.45\textwidth]{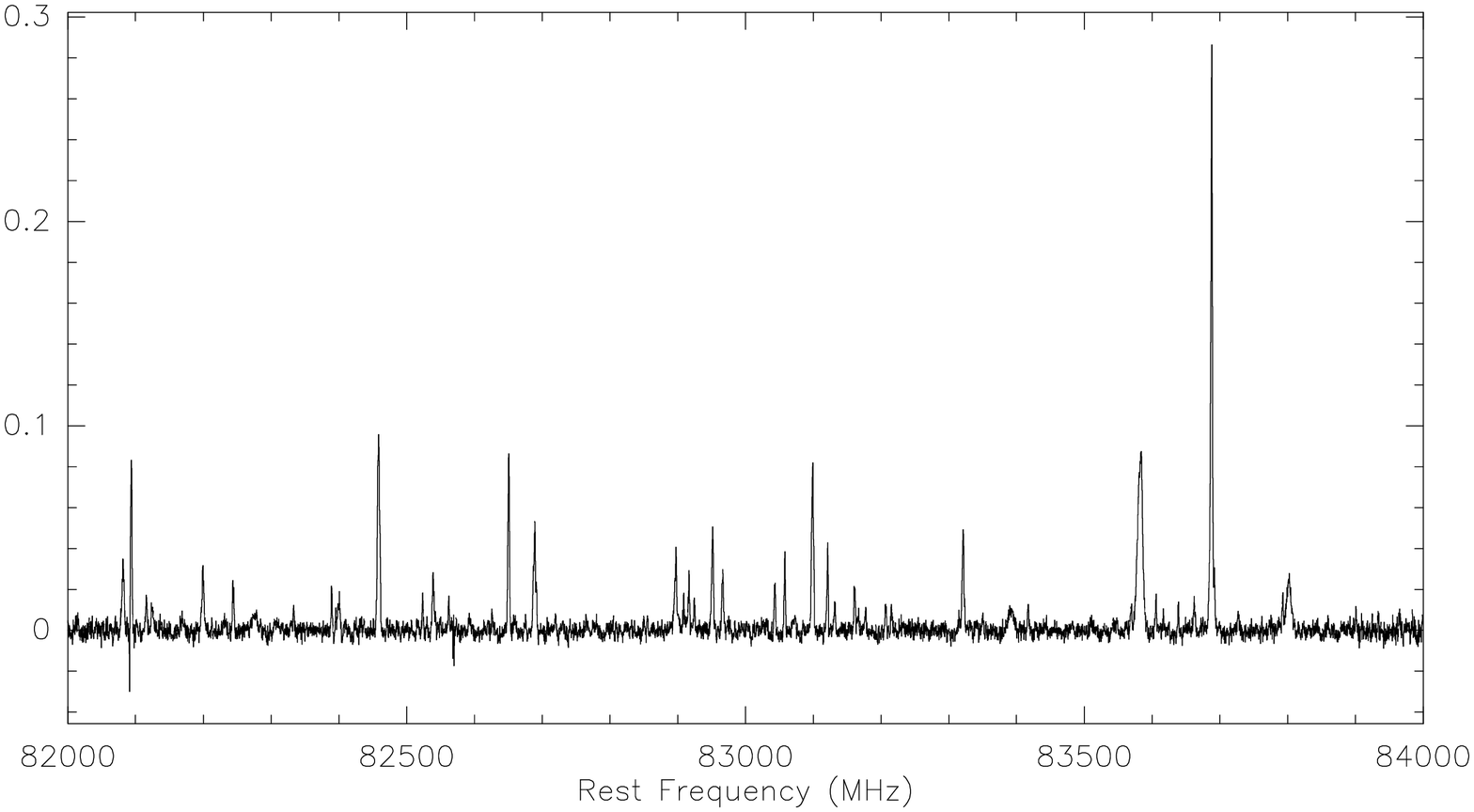}

\includegraphics[width=0.45\textwidth]{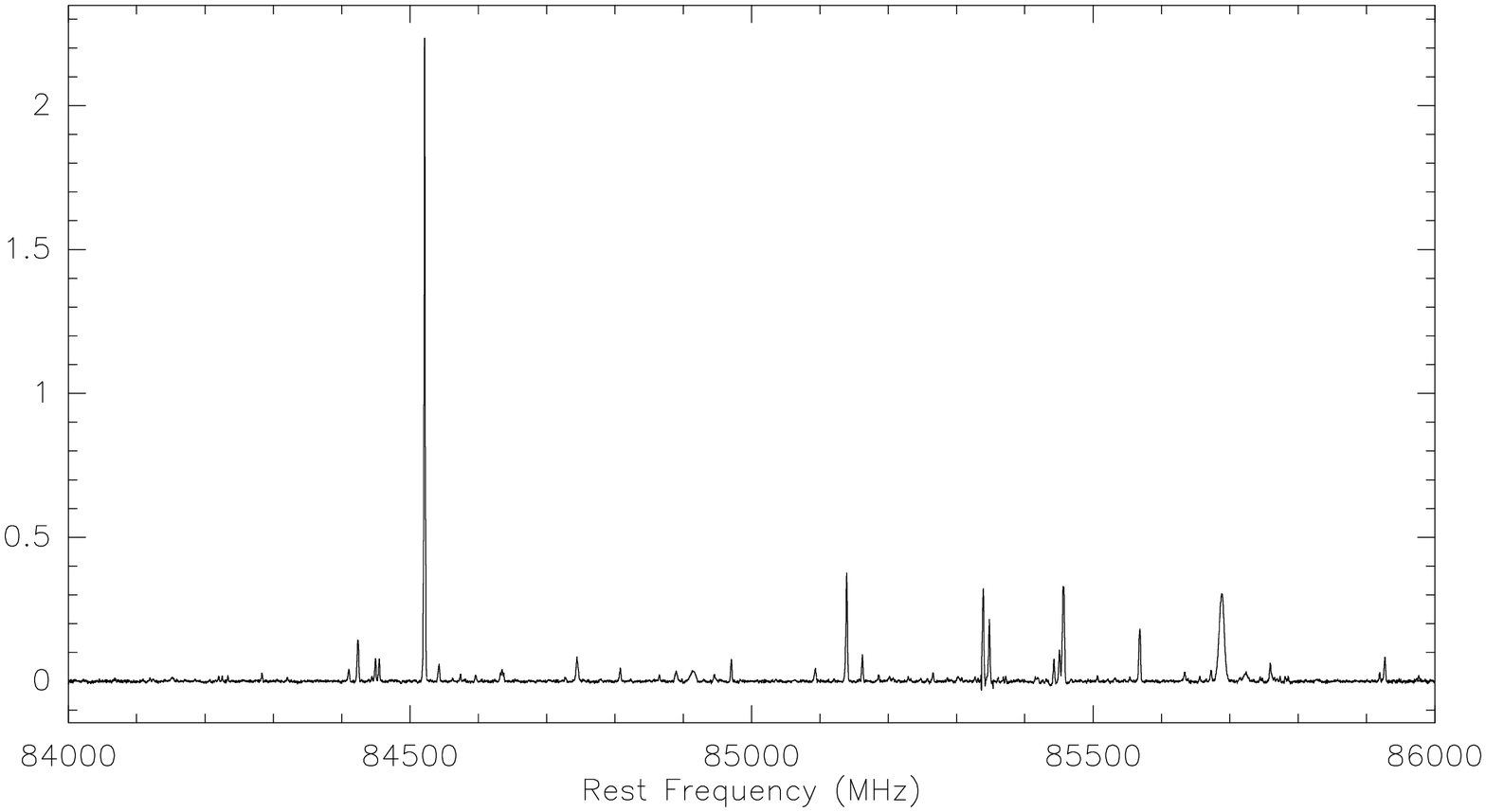}
\includegraphics[width=0.45\textwidth]{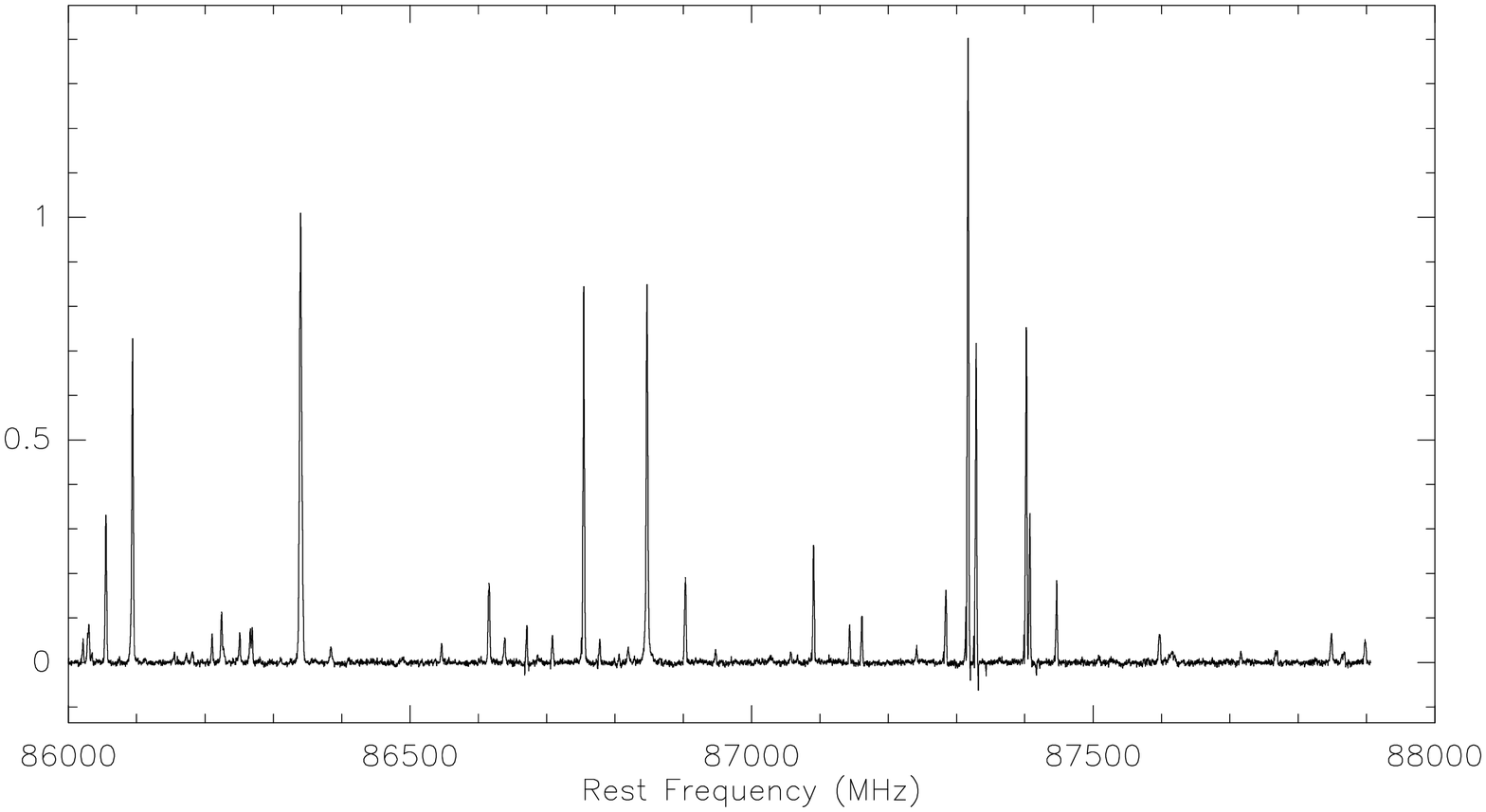}
\caption{W51e1/e2 spectra in the range of frequencies 68--88 GHz}
\label{spec}
\end{figure}
\end{center}

\section{Results} \label{sec:res}
The source spectrum covering the frequency range 68--88~GHz is shown in Fig.~\ref{spec}. Multiple radio lines of various molecules have been detected, as well as radio recombination lines of hydrogen and helium. The primary identification of the detected lines was carried out using the database of molecular radio lines detected in space~\citep{lovas09}\footnote{http://physics.nist.gov/}. However, many of the detected lines were missing from this database. The identification of these lines was carried out using the WEEDS extension of the CLASS software package. For this purpose, we constructed a model spectrum of W51e1/e2, based on the results of the survey by~\cite{2010ARep...54.1084K}, performed with the same antenna at a close (partly overlapping) frequency range. When constructing the spectrum, the local thermodynamic equilibrium (LTE) case was assumed. The column densities and excitation temperatures of those molecules whose parameters were determined using rotational diagrams were taken from Table 3 in~\cite{2010ARep...54.1084K}, and for those molecules whose column densities were obtained from the integrated intensity of a single line, from Table 4 of the same work. In the data reduction process, detected molecules whose lines did not fall into the spectral range of the survey by~\cite{2010ARep...54.1084K} were added to the model.

Methylcyanoacetylene (CH$_3$C$_3$N) was originally detected using spectral-line stacking. The stacking of a large number of spectral lines of the same molecule makes it possible to significantly increase the signal-to-noise ratio and even detect  molecules whose individual lines are not seen in the noise. This method is described in,~e.g.,~\citet{2021NatAs...5..188L}. Our work using this method is far from being complete, and its results will be published subsequently. However, we have already detected CH$_3$C$_3$N, and added it to the model, and after that we could identify several weak lines of this molecule in the observed spectrum. 

\begin{deluxetable*}{cccclcclc}
\tablenum{1}
\tablecaption{Gaussian parameters of the detected lines.\label{tab:gauss}}
\tablewidth{0pt}
\tablehead{\colhead{Comp.} &
\colhead{Frequency} & \colhead{Molecule} & \colhead{Transition} & \colhead{$\int T_RdV$} & \colhead{$V_{LSR}$} & \colhead{$\Delta V$} & \colhead{$T_R$}&\colhead{Notes$^{a,b,c}$}\\
\colhead{}&\colhead{(MHz)} &\colhead{} &\colhead{}& (K km s$^{-1})$ & \colhead{(km s$^{-1})$} & \colhead{(km s$^{-1}$)} & \colhead{(K)} & \colhead{()}
}
\decimalcolnumbers
\startdata
0&70534.033 & H$^{13}$CCCN      & $8-7$                  &0.36(0.04)&  56.55(0.41)  &   9.09(0.91)  & 0.036 & \\ 
0&70583.25  & RRL               & H 70$\delta$           &2.32(0.06)&  58.68(0.40)  &  32.39(0.86)  & 0.06 & \\ 
0&70650     & U                 &                        &0.12(0.03)&  56.87(0.92)  &   6.81(1.81)  & 0.016 &  new; m\\
0&70678.633 & CH$_2$            &$4_{0,4}-3_{1,3}\;F=3-2$&0.42(0.04)&  58.20(0.35)  &  11.00(0.88)  & 0.034 & \\ 
0&70679.543 & CH$_2$            &$4_{0,4}-3_{1,3}\;F=4-3$&0.50(0.04)&  58.20(0.35)  &  11.00(0.88)  & 0.044 & \\ 
0&70680.720 & CH$_2$            &$4_{0,4}-3_{1,3}\;F=5-4$&0.62(0.04)&  58.20(0.35)  &  11.00(0.88)  & 0.054 & \\ 
0&70784.46  & RRL               & H 75$\epsilon$         &1.26(0.10)&  60.12(1.13)  &  27.94(2.25)  & 0.04 &  \\ 
0&70947     & U                 & -------                &0.22(0.04)&  53.99(0.76)  &   8.40(1.67)  & 0.024 & new\\ 
0&70976.795 & t-CH$_3$CH$_2$OH  & $5_{2,3}-5_{1,4}$      &0.38(0.04)&  55.58(0.50)  &   8.77(1.14)  & 0.042 & bl.\\ 
0&70979.627 & C$_2$H$_5$CN    & $8_{0,8}-7_{0,7}$      &0.28(0.04)&  57.36(0.56)  &   7.55(1.38)  & 0.034 & bl.\\ 
0&71024.781 & H$_2^{13}$CO      & $1_{0,1}-0_{0,0}$      &1.02(0.04)&  56.27(0.14)  &   8.28(0.34)  & 0.12 &  \\ 
0&71062.67  & RRL               & H 56$\beta$            &8.16(0.10)&  58.29(0.18)  &  31.72(0.43)  & 0.242 & \\ 
0&71091.63  & RRL               & He 56$\beta$           &0.70(0.06)&  57.13(0.96)  &  23.34(2.43)  & 0.028 & \\ 
0&71155.210 & $^{13}$CH$_3$OH   & $1_1-2_0E$             &0.12(0.02)&  54.57(0.35)  &   3.81(0.67)  & 0.028 & new\\
0&71192.938 & OC$^{34}$S        & $6-5$                  &0.34(0.04)&  56.64(0.28)  &   6.25(0.65)  & 0.05 & new\\ 
0&71304.063 & HCOOCH$_3$        & $17_{4,13}-17_{3,14}E$ &0.22(0.04)&  60.13(0.41)  &   6.51(1.25)  & 0.03 &  new\\ 
0&71309.952 & HCOOCH$_3$        & $15_{3,12}-15_{2,13}E$ &0.44(0.04)&  54.98(0.44)  &  10.09(1.01)  & 0.04 & \\ 
0&71325     & B8                &                        &0.20(0.02)&  59.47(2.70)  &   7.45(0.54)  & 0.026 & \\ 
0&71350     & B6                &                        &0.16(0.02)&  57.59(0.43)  &   5.48(0.98)  & 0.026 &  new \\ 
0&71357.83  & RRL               & H 83$\eta$             &0.68(0.04)&  62.61(0.85)  &   23.60(1.71) & 0.02 & \\ 
1&71464.138  &$^{13}$CH$_3$CN & $4_1-3_1$              &0.14(0.024)& 53.08(0.78)  &   8.85(2.16)  & 0.01 & m\\ 
2&71465.520   &$^{13}$CH$_3$CN & $4_0-3_0$              &          &               &               &         & \\ 
0&71500.528 & C$_2$H$_5$CN    & $8_{2,7}-7_{2,6}$      &0.64(0.04)&  58.90(0.28)  &  11.38(0.65)  & 0.052 & 
\enddata
\tablecomments{Table 1 is published in its entirety in the machine-readable format. A portion is shown here for guidance regarding its form and content. Comp means component of blended lines. Spectral features consisting of two or     more unresolved lines are flagged with Comp $> 0$ and appear consecutively with the primary line used for the radial velocity listed first. 0 = not blended; 1 = the line that is the source of radial velocity in a blended spectral feature; 2 = other line in a blended spectral feature.\\
$^a$ ''new'' means that the line is absent from the ''List of Recommended Rest Frequencies for Observed Interstellar Molecular Microwave Transitions''~\citep{lovas09} \\
$^b$ ''m'' means marginal detection \\ 
$^c$ ''bl.'' means that the line is blended with the neighboring line, but the lines are resolved}
\end{deluxetable*}

Often, the newly discovered spectral features turned out to be blends of lines of a large number ($>4$) of the various molecules included in the model spectrum. When it was not possible to separate the contributions of these lines, we evaluated the parameters of the entire blend, and instead of writing the molecule name, we wrote Bx in the second column of Table~\ref{tab:gauss},  where ''x'' is the number of lines\footnote{We did the same when a specific line is given in the Lovas database at the frequency of the detected spectral feature, but the modeling shows that actually this feature is a blend of lines of several molecules}. At the same time, lines with an upper-level energy above 2000~K and lines with the value of the Einstein A-coefficient being three or more orders of magnitude lower than that of the strongest line were not taken into account. 

In total, 79 molecules and their isotopic species were detected at 4~mm, beginning from simple diatomic and triatomic molecules, such as SO, SiO, and CCH, to complex organic compounds, such as CH$_3$CN, CH$_3$OCH$_3$, or CH$_3$COCH$_3$. 
The list of detected molecules is given in Table~\ref{tab:detmol}.

A significant portion of the obtained results qualitatively repeats the results of the 3~mm survey. A notable portion of all the detected molecules are those that are typical for hot cores. In particular, we found lines of neutral molecules like  HCOOCH$_3$, CH$_3$OCH$_3$, CH$_3$CH$_2$OH, and CH$_3$COCH$_3$, which, according to modern concepts, exist in the gas phase only in hot cores and in hot postshock gas. Another consequence of the fact that we were observing hot gas was the detection of lines from the vibrationally excited levels of, e.g., C$_4$H and HC$_3$N, with upper-level temperatures\footnote{Upper-level energies divided by the Boltzmann constant} on the order of several hundred Kelvins. Such lines are likely to be detectable only if they arise from hot gas with a temperature on the order of 100 K or higher.

In addition to the lines of neutral molecules, lines of various molecular ions were found. Some of them (HC$^{18}$O$^{+}$, HC$^{17}$O$^+$, H$^{13}$CO$^+$, DCO$^+$, and HCS$^+$) are usually present in molecular clouds with a large absorption in the visible spectral range $A_V$.

We could not detect any line of MgCN or NaCN, molecules that were tentatively found by~\citet{2010ARep...54.1084K}. Therefore, we believe that the tentative detection of these molecules by~\citet{2010ARep...54.1084K} was a result of the incorrect identification of lines wrongly attributed to these molecules. Also, we could not confirm the possible detection of four molecules, HOCH$_2$COOH, H$_2$SO$_4$, GG'a-CH$_3$CHOHCH$_2$OH, and l-C$_7$H$_2$, made at 3~mm using spectral-line stacking.

\begin{table}
\tablenum{2}
\caption{Molecules Detected in W51e1/e2 in the 4 mm Wavelength Range.}
\label{tab:detmol}
\begin{tabular}{|l|l|}
\noalign{\medskip}
\hline\noalign{\smallskip}
Diatomic& NS, Si$^{18}$O, SiO, $^{29}$SiO, $^{30}$SiO?, SO, $^{34}$SO;\\
Triatomic& CCH, $^{13}$CCH, C$^{13}$CH, CCS, DCO$^+$, DCN, DNC, HC$^{15}$N, H$^{13}$C$^{15}$N, HCO, HDO, HNO?, H$^{13}$CO$^+$, \\ 
 & HC$^{17}$O$^+$, HC$^{18}$O$^+$, HCS$^+$, H$^{13}$CN, HC$^{15}$N, HN$^{13}$C, N$_2$O, OCS, O$^{13}$CS, OC$^{34}$S, SO$_2$, $^{34}$SO$_2$\\ 
Four-atom&
C$_3$N,  C$_3$S, c-$^{13}$CCCH, HNCO, l-C$_3$H, H$_2$CO, 
H$_2$CS, H$_2$C$^{34}$S?, HOCO$^+$, NH$_2$D\\ 

Five-atom&
c-C$_3$H$_2$, C$_4$H, CH$_2$CO, H$^{13}$CCCN, HC$^{13}$CCN, HCC$^{13}$CN, HCCC$^{15}$N, HCOOH, c-CC$^{13}$CH$_2$, H$_2$CCN?, \\
&HC$_3$N \\
Six-atom&
$^{13}$CH$_3$OH, CH$_3$CN, $^{13}$CH$_3$CN, CH$_3$$^{13}$CN?,
CH$_3$$^{18}$OH, CH$_3$OH, CH$_3$SH, NH$_2$CHO\\  
Seven-atom&
C$_6$H, C$_2$H$_3$CN, CH$_3$CHO, c-C$_2$H$_4$O, CH$_3$CCH, 
CH$_3$NH$_2$, HC$_5$N\\  
Eight-atom&
CH$_3$C$_3$N, HCOOCH$_3$, H$_2$C$_3$HCN?, HCOCH$_2$OH\\ 
Nine-atom&
CH$_3$CH$_2$OH, CH$_3$OCH$_3$, CH$_3$OHCHO, C$_2$H$_5$CN\\ 

Ten-atom&
aGg'-HOCH$_2$CH$_2$OH, CH$_3$COCH$_3$\\

\hline
\end{tabular}
\end{table}

\subsection{Rotation diagrams}
\label{subsec:rotdiagr}
\begin{figure*}
\begin{center}
\includegraphics[width=0.8\textwidth]{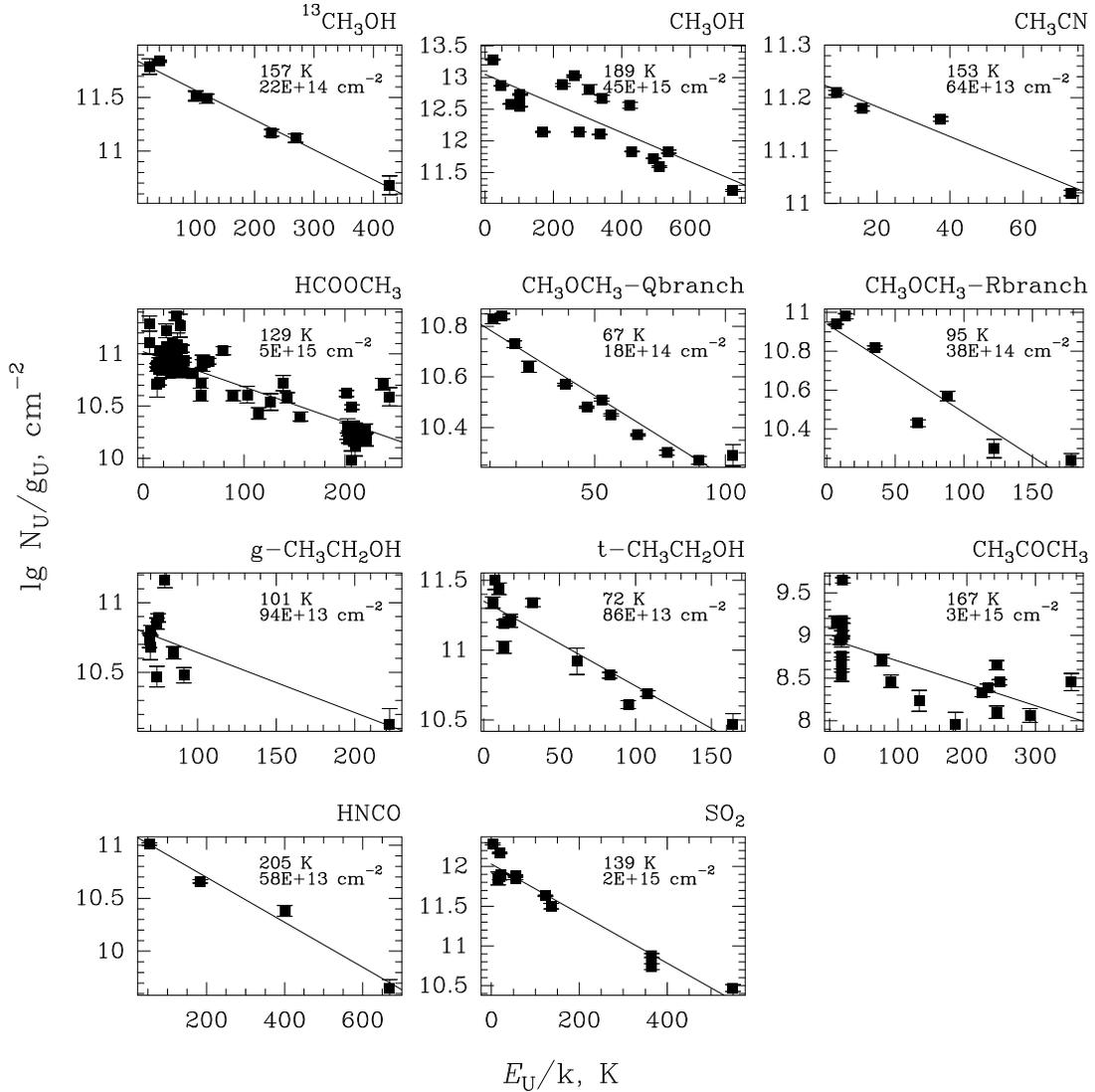}
\caption{RDs that yield $T_{\rm rot}$ about 100~K or higher.}
\label{rd1}
\end{center}
\end{figure*}

\begin{figure*}
\begin{center}
\includegraphics[width=0.8\textwidth]{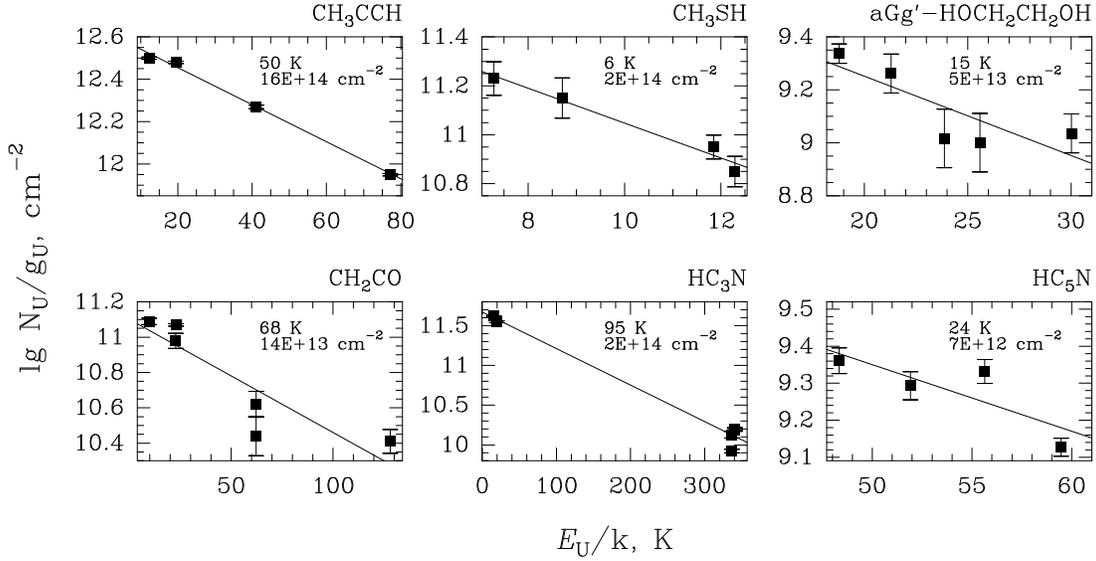}
\caption{RDs that yield $T_{\rm rot}$ much below 100~K. In addition, we show the HC$_3$N RD ($T_{rot}=95$~K) here, as it is jointly discussed with the HC$_5$N RD.}
\label{rd2}
\end{center}
\end{figure*}

\begin{figure*}
\begin{center}
\includegraphics[width=0.8\textwidth]{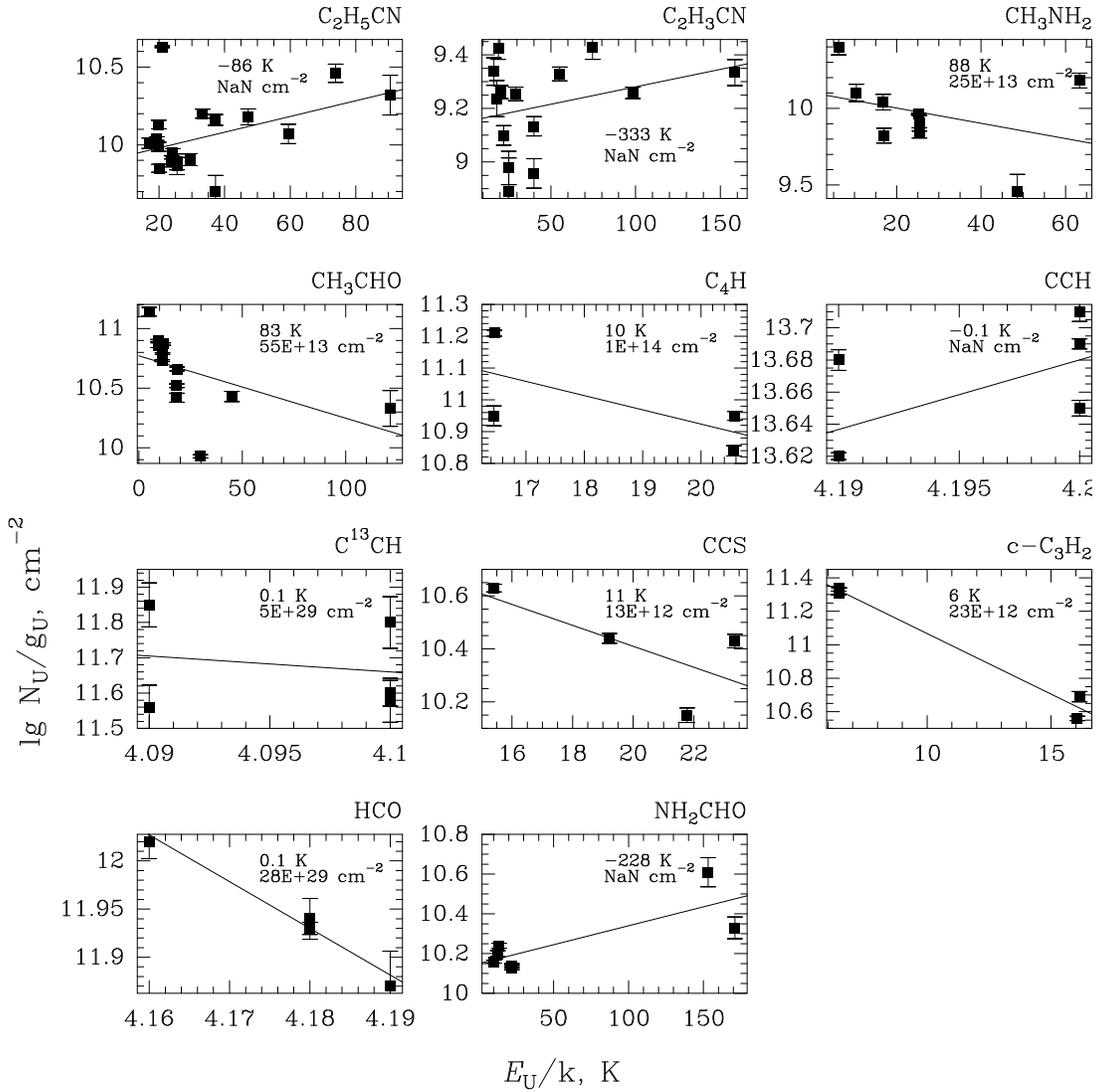}
\caption{"Peculiar" RDs}
\label{rd3}
\end{center}
\end{figure*}

For all molecules with four or more spectral lines detected, we have constructed rotation diagrams (RDs) to estimate the rotational temperature and column density of the molecule. These diagrams are shown in Figs.~\ref{rd1}--\ref{rd3}, and the determined parameters with errors are presented in Table~\ref{tab:rottab}. For several molecules (shown in Fig.~\ref{rd1}), there is a clear correlation between the level population and energy, and the rotational temperature is on the order of 100--150~K, roughly in agreement with the temperatures of the hot cores e2 and e8, obtained by~\cite{1998ApJ...494..636Z,2004ApJ...606..917R}, or~\cite{2016A&A...589A..44G}. The significant scatterings of the points in some of these diagrams could be caused by the deviations of the energy-level populations from the LTE, the heterogeneity of the regions in which the corresponding molecules emit, and/or the high optical depths of some lines. However, it is clear that all these molecules emit mainly in the hot cores e2 and e8. 

We attribute the molecules CH$_3$OCH$_3$ and CH$_3$CH$_2$OH to the same group. However, for dimethylether (CH$_3$OCH$_3$), the RD constructed using the lines $\Delta J=0$ (Q-branch) yields $T_{\rm rot}$ of 67~K, and only the RD constructed using the lines $\Delta J=1$ (R-branch) yields $T_{\rm rot}$ equal to 104~K. This meaningful difference shows that in the case of CH$_3$OCH$_3$ the deviations of the level populations from the LTE are notable, in which case it is difficult to estimate the source kinetic temperature. The relatively moderate rotational temperature of CH$_3$OCH$_3$, with Q-branch transitions, compared to that of the other hot core molecules in W51e1/e2, may mean that  dimethylether sources go beyond the boundaries of hot cores with temperatures $\gtrsim 100$~K and comprise colder regions more distant from protostars. Note that dimethylether has already been found in even the cold gas phase in the prestellar core L1689B~\citep{2012A&A...541L..12B}. In the case of ethanol (CH$_3$CH$_2$OH), the rotational temperature of the gauche conformer is on the order of 100~K, but Fig.~\ref{rd1} shows that it strongly depends on a single point and, therefore, is determined very unreliably (see Table~\ref{tab:rottab}). The rotational temperature of the trans conformer is determined much more reliably, and constitutes only 72~K. Therefore, it is possible that ethyl alcohol sources also go beyond the boundaries of hot cores.

The symmetric-top molecule methylacetylene (CH$_3$CCH) is a good thermometer for interstellar gas, as the rotational temperature obtained from the lines $J_K-(J-1)_K$ is close to the gas kinetic temperature~\citep{1984A&A...130..311A}. Interferometric observations of the low-mass protostellar binary IRAS 16293-2422 show that the emission of methylacetylene arises in the hot corinos IRAS 16293A and IRAS 16293B, and that the rotational temperature obtained from the CH$_3$CCH lines is on the order of 100~K~\citep{2019A&A...631A.137C}. An even higher CH$_3$CCH temperature, 320~K, has been reported as a result of Atacama Large Millimeter/submillimeter Array observations of the massive star formation region AFGL4176~\citep{2019A&A...628A...2B}. However, the rotational temperatures obtained from the CH$_3$CCH observations with single-dish antennas have proved to be significantly lower, within the range 20--50~K~\citep{1984A&A...130..311A, 2002ARep...46..551A}. This fact shows that in addition to the hot cores, methylacetylene exists in a colder gas, and at scales significantly larger than the sizes of hot cores, the contributions of cold regions dominate the radiation of methylacetylene. We obtained a methylacetylene rotational temperature of only~49~K~(see~Fig.~\ref{rd2}). Such a low value definitely indicates the presence of colder gas in the observed region and the existence of methylacetylene in this gas.

As for ketene (CH$_2$CO), its rotational temperature is also fairly low, 68~K. This molecule, in addition to the hot regions, has been observed in cold gas~\citep{1989ApJ...342..871I, 2012A&A...541L..12B}, so we believe that the low rotational temperature could appear due to the contribution of cold gas.

The RD for cyanoacetylene (HC$_3$N) includes both low-energy levels in the ground vibrational state and high-energy levels in the excited vibrational states. The rotational temperature proved to be 95~K. The RD for the next member of the cyanopolyyne sequence, cyanodiacetylene (HC$_5$N), includes only low-energy levels in the ground vibrational state, and the rotational temperature proved to be only 24~K. The difference in the rotational temperatures of the two related molecules can be explained as follows. Suppose that cyanopolyynes are present in both hot cores and cold gas. The transitions between the HC$_3$N levels in the ground vibrational state are about 50 times stronger than the same HC$_5$N transitions (see Table~\ref{tab:gauss}). Since the transitions between the HC$_3$N levels in excited vibrational states are very weak, it is not surprising that such HC$_5$N transitions were not detected due to a lack of sensitivity. Ground-state cyanopolyyne lines arise mostly in a cold gas, and excited-state lines in hot gas. Therefore, the RD of HC$_5$N refers to a cold gas and indicates the presence of a gas with a kinetic temperature of about 24 K. As for HC$_3$N, this RD is related to a mixture of a cold gas with $T_{kin}$ about 24~K and a hot gas in e2 and e8. The diagram yields some averaged value of $T_{\rm rot}$. If we could remove the contribution of the cold gas, the points on the left side of the diagram, which correspond to the lines in the ground state, would go down, and the rotational temperature would increase and would probably become close to the rotational temperatures of SO$_2$ and CH$_3$OH, etc.

We believe that for the molecules listed above, RDs yield fairly reasonable rotational temperatures and column densities. However, there are a number of other molecules with more than four lines detected, which are sufficient to construct RDs, but the parameters obtained with these diagrams hardly coincide even approximately with the temperatures of the emitting regions and the molecular column densities. 

The RD based on the lines of methyl mercaptan (CH$_3$SH) shows a small scatter of points, but the rotational temperature turned out to be extremely low, only 6~K (see~Fig.~\ref{rd2}). It is unlikely that the methyl mercaptan source is so cold, as such a low temperature is unusual even for IRDCs, whose temperatures vary between 10 and 20~K~\citep[e.g.,][]{2006A&A...447..929P}, and is extremely unlikely for regions of high-mass star formation at later stages of evolution, such as W51e1/e2. We believe that this temperature can be explained as follows.

The structure of the methyl mercaptan molecule is similar to the structure of the methanol molecule, and the system of the rotational levels and transitions of CH$_3$SH should be similar to that of methanol. When the gas number density is lower than $\sim 10^7$~cm$^{-3}$, the populations of the rotational levels of methanol differ markedly from the LTE. Based on the modeling of methanol emission by the Large Velocity Gradient method, \citet{2016ARep...60..702K} analyzed the use of methods based on the LTE assumption---particularly RDs---in the analysis of methanol observations. In particular, they considered the lines $3_K-2_K$; only $3_K-2_K$ methyl mercaptan lines were detected in our survey. 
According to~\cite{2016ARep...60..702K}, when the gas number density is insufficient to provide the LTE, the RDs constructed using such lines yield rotational temperatures that are much lower than the kinetic temperature. Moreover, such rotational temperatures depend on the gas number density, rather than on the temperature. We believe that the low $T_{\rm rot}$ value for methyl mercaptan appears for the same reason. Note that in our survey we observed completely different lines of CH$_3$OH and $^{13}$CH$_3$OH, which explains the difference in the methanol and methyl mercaptan rotational temperatures. 

Anyway, the RD for methyl mercaptan clearly shows that even a low scatter of points on an RD and small formal errors of $T_{\rm rot}$ and molecular column density do not guarantee that the obtained parameter values represent the source parameters. This happens, in particular, because the LTE condition is often violated in the interstellar medium (ISM). The statement is especially important if a small series of similar lines are used to construct an RD. The safe usage of this method is only possible after careful analysis using statistical equilibrium calculations, such as the one that was carried out for methanol by~\citet{2016ARep...60..702K} or the one for methylacetylene by~\citet{1984A&A...130..311A}. Otherwise, the results obtained with RDs can only be trusted if they are supported by other results, which, in the case of well-studied sources such as W51e1/e2, means that they can be trusted with caution only if the results are consistent with the source parameters and 
the chemistry of the explored molecules.

\begin{figure}
\begin{center}
\includegraphics[width=0.5\textwidth]{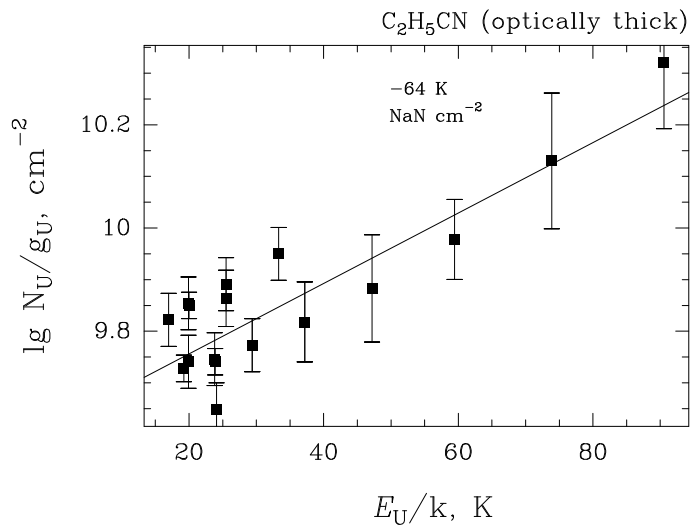}
\caption{The RD build from optically thick lines of ethylcyanide.}
\label{optthick}
\end{center}
\end{figure}

 A low rotational temperature was obtained for ethylene glycol (aGg'-HOCH$_2$CH$_2$OH; 15~K). Ethylene glycol in the gas phase in the ISM has been found only in hot cores and their counterparts in regions of low-mass star formation (hot corinos); in addition, it has been detected in comets \citep[e.g.][and references therein]{2018MNRAS.475.2016C}. Therefore, we believe that the low rotational temperature is hardly caused by a contribution of cold gas. Rather it can be explained by deviations from the LTE. 
 
\begin{deluxetable}{lcr}
\tablenum{3}
\tablecaption{Rotational Temperatures and Beam-averaged Column Densities of the Detected Molecules Determined Using RDs\label{tab:rottab}}
\tablewidth{0pt}
\tablehead{
\colhead{Molecule} & \colhead{$T_{\rm rot}$} & \colhead{$N_{mol}$}\\
\colhead{} &\colhead{(K)} &\colhead{($10^{13}$~cm$^{-2}$)}}
\startdata
$^{13}$CH$_3$OH    & 157(8)      & 220(13)  \\
CH$_3$OH           & 189(32)     & 4500(1700) \\
CH$_3$CN           & 153(20)     & 64(2)\\
HCOOCH$_3$         & 129(9)      & 500(33)\\
CH$_3$OCH$_3$(Q)   & 67(5)       & 180(14)\\
CH$_3$OCH$_3$(R)   & 95(16)      & 380(64)\\
$g-$CH$_3$CH$_2$OH & 101(33)     & 94(36)\\
$t-$CH$_3$CH$_2$OH & 72(9)       & 86(11)\\ 
CH$_3$COCH$_3$     & 167(37)     & 300(70)\\
HNCO               & 205(17)     & 58(10)\\
SO$_2$             & 139(10)     & 200(25)\\
CH$_3$CCH          & 50(2)       & 160(6) \\
CH$_2$CO           & 68(16)      & 14(4) \\  
HC$_3$N            & 95(6)       & 20(4)\\ 
HC$_5$N            & 24(10)      & 0.7(0.9)\\ 
\enddata
\tablecomments{The values in parentheses are $1\sigma$ errors}
\end{deluxetable}

We built RDs for 11 more molecules (Fig.~\ref{rd3}), but failed to draw conclusions based on them. Most of these diagrams look peculiar and demonstrate strong scatterings of points. In the cases of HCO, CCH, $^{13}$CCH, and C$_4$H, the ranges of level energies are extremely small (for the first three of these molecules, we observed different hyperfine components of the same rotational transitions); therefore, even small deviations of the level populations from the LTE can lead to significant changes in the slopes of the approximating straight lines and, therefore, strongly distort the rotational temperature. 

The rotational temperatures for C$_2$H$_5$CN, C$_2$H$_3$CN, NH$_2$CHO, and CCH proved to be negative. The RD analysis works well for optically thin emission (and for a homogeneous source). If some of the transitions used are optically thick, they will deviate from a straight line in the RD. In such cases, the RD analysis may result in peculiar values for $T_{\rm rot}$ and N, if they are not compensated for. \citet{gins16} found that the optical depths of the ammonia lines seen in absorption against the HC \hii region W51e2-W vary in the range $10 - 100$. Therefore, it is necessary to understand whether the lines found in this survey are optically thin.

Obviously, a line with $T_R \lesssim 1$~K will be optically thin when the source size is on the order of or larger than the beam size. However, the emission of many molecules arises in hot cores, whose angular sizes in W51e1/e2 are only about 2 arcsec (see Figs.~5 and 8 in~\citet{gins16}). One can estimate at what $T_R$ values the lines will be optically thin as follows. 

If the lines appear in a thermalized gas with $T_{rot} \sim 150$~K, the brightness temperature of a line with optical depth $\tau = 1$ will be 95~K. Assuming that the size of the hot cores is about $2''$, we find that the peak main-beam brightness temperature $T_R$ of such a line will be about 0.3. Most of the lines from Tab.~\ref{tab:gauss} are significantly weaker, and hence they may be optically thin. This conclusion is supported by the ratios of isotopologue abundances presented in Table~\ref{tab:ratios}. The ratios were calculated from the column densities presented in Tables~\ref{tab:rottab} and \ref{tab:coldentab}. However, in cases where the rotational temperature was not known, and/or the observed lines of the primary and secondary isotopologues were not the hyperfine components of a single rotational transition, only the lines detected in both the main and the secondary isotopologues were used.
The comparisons with the $^{12}$C/$^{13}$C, $^{16}$O/$^{18}$O, $^{32}$S/$^{34}$S, and $^{28}$Si/$^{29}$Si isotopic ratios show that most of the abundance ratios are consistent with the optical depths of the main isotopologues less than unity. The exceptions are CH$_3$OH (but this molecule has many lines brighter than 0.3~K),  H$_2$CO, and HC$_3$N (but the used lines for these molecules are brighter than 0.3~K). The results on OCS are ambiguous: the optical depths of the lines are less than unity, according to the OCS/OC$^{34}$S ratio, but are close to 3, according to the OCS/O$^{13}$CS ratio. Anyway, the brightness temperatures of the OCS lines are higher than 0.3~K. Thus, the ratios of isotopologue abundances are consistent with the assumption that the weak lines ($T_{br}\lesssim 0.3$~K) detected at 4~mm are usually optically thin. 

\begin{deluxetable}{llr}
\tablenum{4}
\tablecaption{Abundance Ratios\label{tab:ratios}}
\tablewidth{0pt}
\tablehead{
\colhead{Isotope} &\colhead{Isotopologue}&\colhead{Abundance}\\
\colhead{Ratio}   &\colhead{Ratio}       &\colhead{Ratio}}
\startdata
\multirow{9}{*}{59$^a$ ($^{12}$C/$^{13}$C)} 
                  &CCH/C$^{13}$CH             & 45\\  
                  &CCH/$^{13}$CCH             & 74\\
                  &CH$_3$CCH/$^{13}$CH$_3$CCH & 43\\ 
                  &CH$_3$CN/CH$_3$$^{13}$CN   & 40\\ 
                  &CH$_3$CN/$^{13}$CH$_3$CN   & 119\\
                  &CH$_3$OH/$^{13}$CH$_3$OH   & 20\\
                  &H$_2$CO/H$_2$$^{13}$CO     & 37$^e$\\  
                  &HC$_3$N (v=0)/H$^{13}$CCCN & 36$^e$\\
                  &HC$_3$N (v=0)/HC$^{13}$CCN & 49$^e$\\
                  &HC$_3$N (v=0)/HCC$^{13}$CN & 34$^e$\\
                  &OCS/O$^{13}$CS             & 22$^e$\\ 
\hline
400$^b$ ($^{16}$O/$^{18}$O)
                  &SiO/Si$^{18}$O             &$>81^e$\\
\hline
\multirow{3}{*}{22$^c$ ($^{32}$S/$^{34}$S)}   
                  &OCS/OC$^{34}$S             & 16$^e$\\
                  &SO/$^{34}$SO               & 16$^e$\\
                  &SO$_2$/$^{34}$SO$_2$       & 15\\
\hline
19$^d$ ($^{28}$Si/$^{29}$Si)
                  &SiO/$^{29}$SiO             & 15$^e$
\enddata
\tablecomments{\\$^a$ calculated using Eq.~5 in \citet{2005ApJ...634.1126M}\\ 
$^b$ see Fig.~2 in \citet{1994ARAA..32..191W} \\
$^c$ \citet{1994ARAA..32..191W}\\ 
$^d$ \citet{2017ApJ...839..123M} \\
$^e$ the abundance ratios were determined using only the transitions detected both in the main and secondary isotopologues}
\end{deluxetable}
 
The lines of any molecule whose RD is presented in Figs.~\ref{rd1}--\ref{rd3} are on average much weaker than the lines of methanol. In most cases, their brightness temperatures are below 0.3~K. Therefore, one can expect that they are optically thin. Note that even the $^{12}$CH$_3$OH RD yields a fairly good estimate of the temperature. Therefore it is reasonably to {\em assume} that the other believable-looking RDs, constructed from weaker lines, also yield reasonable estimates of the temperatures and beam-averaged  column densities. The parameters determined with 15 such RDs are presented in Table~\ref{tab:rottab}. However, since the abundances of molecules in the gas phase depend on temperature, some molecules probably exist only in the hottest central regions of e2 and e8. The angular sizes of these regions may be much less than $2''$, and even the very faint lines presented in Table~\ref{tab:gauss} may be optically thick. For example, if a source size is $1''$, and its temperature is 200~K, the lines with $T_R \approx 0.1$~K have optical depths of the order of unity. Therefore, it is possible that some peculiar RDs become so due to the large optical depths of the used lines. For example, the RDs for C$_2$H$_5$CN and C$_2$H$_3$CN show large scatters of points and yield negative rotational temperatures (Fig.~\ref{rd3}). Fig.~\ref{optthick} shows an RD built using the same C$_2$H$_5$CN lines that were used to construct the RD in Fig.~\ref{rd3}, and the brightness temperatures of all the lines were set the same, corresponding to the case of the high optical depth of all lines\footnote{This simplified approach was used because it is impossible to reliably model ethyl cyanide or vinyl cyanide emission due to the absence of collisional constants.}. The rotational temperature turned out to be negative, and the points show a noticeable scatter, although it is much weaker than the scatter in the corresponding diagram in Fig.~\ref{rd3}. We believe that the C$_2$H$_5$CN emission arises both in cold extended sources and in the hottest central regions of e2 and e8, and that the lines arising in the latter regions are optically thick. In the cold regions, the density can be relatively low, allowing significant deviations from the LTE. This explains the strong scatter of points, corresponding to the low-energy levels. With the increase of the level energy the contribution of the cold gas decreases, and the radiation of hot regions becomes predominant. The optically thick lines from these regions provide the slopes of the approximating straight lines that correspond to negative temperatures.

To check this scenario, one needs high-resolution maps of W51e1/e2 in the lines of C$_2$H$_5$CN and C$_2$H$_3$CN. Therefore, further interferometric observations of this source are highly desirable.

The remaining RDs show strong scatters of points that could be caused by deviations from the LTE, source inhomogeneities, errors in determining the parameters of weak lines, as well as contributions of unrecognized weak spectral features to the brightness of the weak lines of the studied molecules.  The column densities of the 11 molecules demonstrating peculiar RDs were estimated using individual lines (see the following section). 

\subsection{Molecular column densities derived from the integrated intensities of single lines}
\label{subsec:slcd}
If the number of detected lines of a molecule was insufficient to construct an RD, we determined the molecule's column density from the integrated intensity of a single line. We did the same in those cases where the built RDs proved to be unsuitable for determining the column density. As with the RDs, blended lines were used only when it was possible to reliably separate the individual components. When several lines proved to be suitable for determining the column density of a molecule, we determined it from each of these lines and then averaged the results.

The column density of a molecule can be determined from a single line only when the rotational temperature is known in advance. Often, it was possible to make reasonable a~priori assumptions about the value of $T_{\rm rot}$. For example, the rotational temperature of the cyanopolyynes H$^{13}$CCCN, HC$^{13}$CCN, and HCC$^{13}$CN was assumed to be 24~K, equal to the $T_{\rm rot}$ of HC$_5$N. This value was obtained for ''cold'' cyanopolyynes (see~Fig.~\ref{rd2}). For CH$_3^{13}$CN and $^{13}$CH$_3$CN, we used $T_{\rm rot}$=153~K, equal to that obtained for the main isotopologue CH$_3$CN. For sulfur-bearing molecules, a hot core temperature of 150~K was taken as $T_{\rm rot}$, etc. However, for a notable portion of the detected molecules, we could not make reasonable assumptions about the value of $T_{\rm rot}$. In these cases, we did the following.
 
\startlongtable
\begin{deluxetable}{lcr}
\tablenum{5}
\tablecaption{Beam-averaged Column Densities Determined from Individual Lines\label{tab:coldentab}}
\tablewidth{0pt}
\tablehead{
\colhead{Molecule} & \colhead{$T_{\rm rot}$} & \colhead{$N_{mol}$}\\
\colhead{} &\colhead{(K)} &\colhead{($10^{12}$~cm$^{-2}$)}}
\startdata
 CCH                   &   49 & 4689  \\ 
 C$^{13}$CH            &   49 &  104  \\ 
 $^{13}$CCH            &   49 &   63  \\ 
 c-C$_3$H$_2$          &   49 &  131  \\ 
 c-C$_2$H$_4$O         &  150 &  107  \\ 
 c-CC$^{13}$CH$_2$     &   49 &    6  \\ 
 l-C$_3$H              &   49 &   14  \\ 
 C$_4$H                &   49 &  111  \\ 
 CCC$^{13}$CH          &   49 &    9  \\ 
 CCS                   &  150 &   45  \\ 
 CH$_2$                &   49 & 8790  \\ 
 aGg'-HOCH$_2$CH$_2$OH &  150 &  341  \\ 
 $^{13}$CH$_3$CCH      &   49 &   37  \\ 
 CH$_2$CN              &   60 &    5  \\ 
 C$_2$H$_3$CN          &   60 &   25  \\ 
 CH$_3^{13}$CN         &  153 &   16  \\ 
 $^{13}$CH$_3$CN       &  153 &   5.4 \\ 
 CH$_3$NC              &  145 &   10  \\ 
 CH$_2$OHCHO           &   60 &   11  \\ 
 C$_2$H$_5$CN         &   60 &   93  \\ 
 CH$_3$C$_3$N          &   25 & $\lesssim$ 1.6 \\
 CH$_3$NH$_2$          &   60 &  189  \\ 
 CH$_3$CHO             &   60 &  602  \\ 
 CH$_3$SH              &  150 & 1006  \\ 
 DCO$^+$               &   60 &    2  \\ 
 DCN                   &  150 &   11  \\ 
 DNC                   &  150 &   11  \\ 
 HC$_3$N               &   24 &   87  \\
 HC$_3$N, v7=1         &  150 &   99  \\ 
 H$^{13}$CCCN          &   24 &    2  \\ 
 HC$^{13}$CCN          &   24 &    2  \\ 
 HCC$^{13}$CN          &   24 &    3  \\ 
 H$_2$CO               &   60 & 2638  \\ 
 H$_2^{13}$CO          &   60 &  123  \\ 
 H$_2$CS               &   60 &  435  \\
 HCO                   &   60 &  254  \\ 
 HDO                   &   60 &   78  \\ 
 HNO                   &   60 &  457  \\ 
 HC$^{18}$O$^+$        &   60 &    5  \\ 
 HCS$^+$               &  150 &  115  \\ 
 HC$^{15}$N            &   25 &    6  \\ 
 H$^{13}$CN            &   25 &   31  \\ 
 HCOOH                 &  150 &  562  \\ 
 HN$^{13}$C            &   25 &   13  \\ 
 HOCO$^+$              &   49 &    4  \\ 
 N$_2$O                &  150 & 2866  \\ 
 NH$_2$CN              &  150 &   22  \\ 
 NH$_2$CHO             &  150 &  386  \\ 
 NS                    &   60 &  396  \\
 OCS                   &  150 & 2130  \\ 
 OC$^{34}$S            &  150 &  130  \\ 
 O$^{13}$CS            &  150 &   95  \\ 
 SiO                   &  150 &  244  \\ 
 $^{29}$SiO            &  150 &   16  \\ 
 Si$^{18}$O            &  150 &   $<3$\\ 
 SO                    &  150 & 3595  \\ 
 $^{34}$SO             &  150 &  221  \\ 
 $^{34}$SO$_2$         &  136 &  136  \\ 
 C$_5$H$^1$            &  150 &    1  \\
\enddata
\end{deluxetable}

The dependence of column density on temperature is described by the well-known formula \mbox{$N=(N_u/g_u)\cdot Q(T)\cdot exp(E_u/(kT))$}, where the partition function $Q(T)$ is proportional to temperature, in the case of linear molecules, and proportional to $T^{1.5}$, in the case of nonlinear molecules. Based on the results obtained with the RDs, one can assume that the rotational temperatures for different molecules fall within the range $\sim 25-150$~K. Taking $T_{rot}=60$~K, we can underestimate the column density of a nonlinear molecule four times, if the rotational temperature is about 150~K, and overestimate it by a factor of 4 if the rotational temperature is close to 25~K. In the case of linear molecules, the maximum relative error will be 2.5. Therefore, in these cases, we adopt $T_{rot}=60$~K and consider the relative error of the column density to be no more than four.

The results are shown in Table~\ref{tab:coldentab}.

Additional results for molecular column densities and  abundances will be given in a forthcoming paper devoted to a joint analysis of the data collected at 4~mm in this survey and at 3~mm by~\citep{2010ARep...54.1084K}.
\section{Molecule formation mechanisms} \label{sec:molform}
The molecular survey of W51e1/e2 provided an inventory of gas-phase molecules as compiled in Figures~\ref{fig:form1} and~\ref{fig:form2}. It is important not only to detect these molecules via microwave spectroscopy, but also to discuss potential formation pathways, particularly of complex organic molecules (COMs) and hydrocarbons, along with nitriles, from the bottom up~\citep{i,i1}.  These formation routes are discussed in this section, first in the context of laboratory experiments elucidating the synthesis of organic molecules in interstellar ices in cold molecular clouds, followed by sublimation into the gas phase in the hot core stage as studied through surface science experiments (Fig.~\ref{fig:form1}). Thereafter, we discuss the predominant formation of hydrocarbons and their nitriles in the gas phase through bimolecular neutral--neutral reactions (Fig.~\ref{fig:form2}). In cold molecular clouds with typical kinetic temperatures near to 10 K, these processes must be exoergic, barrierless, and all transition states must reside lower than the energy of the separated reactants~\citep{ii}. In detail, the surface science experiments replicate the exposure of interstellar analog ices (H$_2$O, NH$_3$, CH$_4$, CO, COS, H$_2$CO, and CH$_3$OH) condensed on interstellar grains at 10 K, with energetic electrons as proxies of energetic galactic cosmic rays (GCRs) penetrating even deep inside cold molecular clouds~\citep{iii}. Once formed inside the ices at 10 K, the transition from the cold molecular cloud to a star-forming region leads to the sublimation of the newly formed molecules from the ices into the gas phase, driven by temperature increases of the ices by up to 300 K. The gas-phase studies exploit crossed molecular beam experiments where the reactants collide under single-collision conditions; this approach guarantees that only primary reaction products are identified~\citep{add3}. To present versatile reaction mechanisms, it is important to rearrange the astronomical inventory of the molecules detected toward W51e1/e2 (Table~\ref{tab:detmol}; Figs.~\ref{fig:form1} and \ref{fig:form2}) according to their chemical functional groups, since the synthesis of molecules with identical functional groups is driven by identical and hence versatile reaction mechanisms~\citep{xxvi3,add1,add2,ii,add3,i1,i}. In Section~\ref{subsec:formation1}, we discuss laboratory experiments exploiting isomer-selective, tunable photoionization; these studies have revealed that (COMs), as compiled in Fig~\ref{fig:form1} can be formed within interstellar ices at 10~K, followed by the sublimation of the new species into the gas phase when the cold clouds transits to a star forming region.  Thereafter, Section~\ref{subsec:formation2} provides potential reaction pathways extracted from molecular beam experiments for hydrocarbons and their nitriles in the gas phase.

\subsection{Formation routes in interstellar ices}  \label{subsec:formation1}

Alcohols. Let us start with the documented reaction pathways of alcohols -- molecules carrying the R-OH functional group with R being a hydrocarbon sidechain -- and their ether isomers -- molecules holding an R-O-R moiety. The reaction pathways to ethanol (CH$_3$CH$_2$OH) and dimethyl­ether (CH$_3$OCH$_3$) were elucidated by Bergantini et al. in methanol (CH$_3$OH)-–methane (CH$_4$) as well as water (H$_2$O)--methane (CH$_4$) ices~\citep{v,2017ApJ...841...96B}. In detail, methane and methanol can interact with GCR proxies to undergo decomposition via reactions (1) and (2), as well as (3) and (4), respectively. These pathways are endoergic from 4.1 to 5.1 eV, with the reaction endoergicity compensated for by some of the kinetic energy of the impinging GCR proxies, which leads to the methyl radical CH$_3$ (1) and the singlet carbene CH$_2$(a$^1$A$_1$) (2), as well as hydroxymethanol CH$_2$OH (3) and methoxy CH$_3$O (4). The exoergic formation of ethanol (CH$_3$CH$_2$OH) and dimethylether (CH$_3$OCH$_3$) was found to be driven mainly by the radical-–radical recombination pathways (5) and (6) and to the less prominent carbene insertion pathways (7) and (8).  Likewise, these studies also revealed the formation of three C$_2$H$_6$O$_2$ isomers, i.e.  ethylene glycol (HOCH$_2$CH$_2$OH), dimethyl peroxide (CH$_3$OOCH$_3$), and methoxymethanol (CH$_3$OCH$_2$OH), via radical--radical recombination (9)-–(11), of which only the diol ethylene glycol (HOCH$_2$CH$_2$OH) was observed in the present study. 

\begin{eqnarray}
\rm CH_4 \rightarrow CH_3\bullet + H\bullet	\\ 
\rm CH_4 \rightarrow CH_2(a^1A_1) + H_2 \\
\rm CH_3OH \rightarrow \bullet CH_2OH + H\bullet \\
\rm CH_3OH \rightarrow CH_3O\bullet + H\bullet \\
\rm \bullet CH_2OH + \bullet CH_3 \rightarrow CH_3CH_2OH\\		
\rm CH_3O\bullet + \bullet CH_3 \rightarrow CH_3OCH_3\\	
\rm CH_3OH + CH_2 \rightarrow CH_3CH_2OH \\
\rm CH_3OH + CH_2 \rightarrow CH_3OCH_3 \\
\rm \bullet CH_2OH + \bullet CH_2OH \rightarrow HOCH_2CH_2OH \\
\rm CH_3O\bullet + CH_3O\bullet \rightarrow CH_3OOCH_3 \\
\rm CH_3O\bullet + \bullet CH_2OH \rightarrow CH_3OCH_2OH
\end{eqnarray} 	

\begin{figure*}
\includegraphics[width=\textwidth]{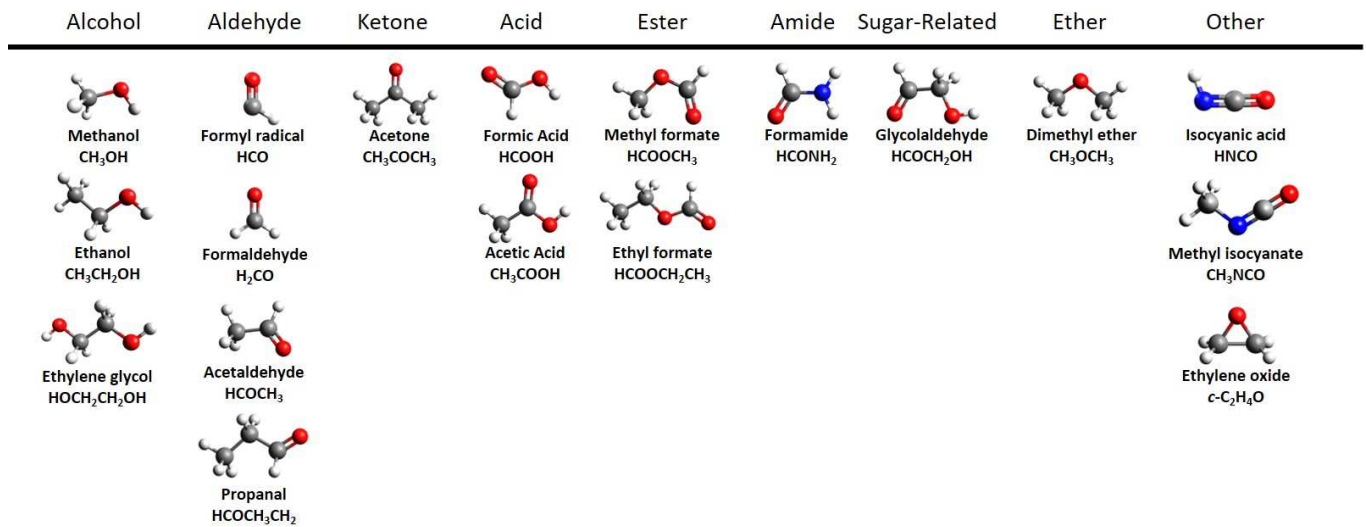}
\caption{The key COMs detected in the present survey, arranged according to their functional groups.}
\label{fig:form1}
\end{figure*}

\begin{figure}[b]
\includegraphics[width=0.8\textwidth]{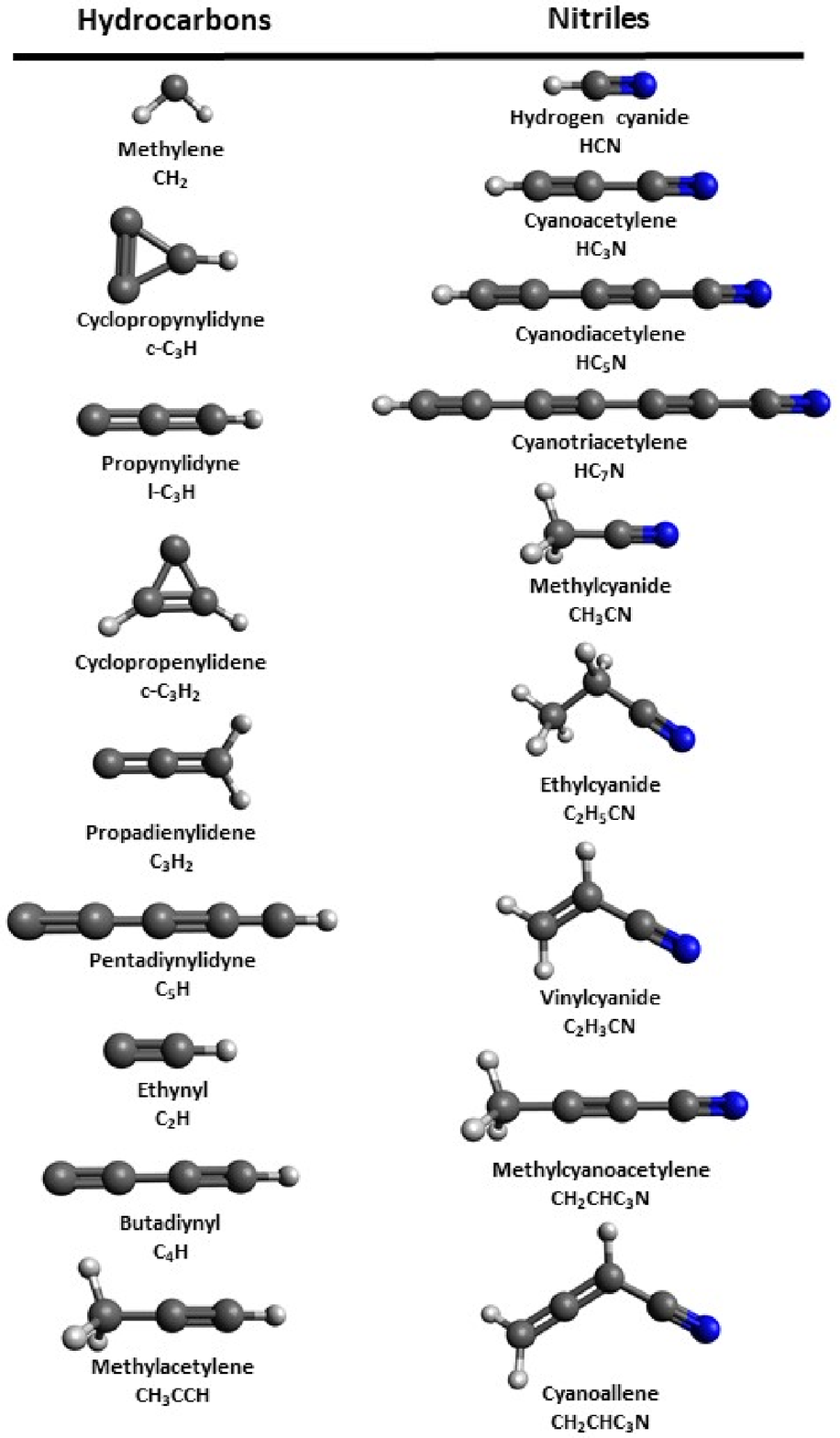}
\caption{The key hydrocarbon molecules along with the closed-shell nitriles detected in the present survey, arranged according to their functional groups}
\label{fig:form2}
\end{figure}

Aldehydes. Aldehydes are defined as organic compounds carrying the formyl moiety (-CHO). In this survey, the three simplest aldehydes---formaldehyde, also known as methanal (H$_2$CO); acetaldehyde, also called ethanal (CH$_3$CHO); and propanal (C$_2$H$_5$HCO)---were detected toward W51e1/e2. Since formaldehyde represents an original component of the interstellar ice inventory (Fig.~\ref{fig:form1}), we only present pathways to acetaldehyde and propanal. As extracted from astrophysical ice analog mixtures of carbon monoxide (CO)--methane (CH$_4$) and carbon monoxide (CO)-– ethane (C$_2$H$_6$), aldehydes can be formed via the exoergic radical--radical recombination of the formyl radical (HCO$\bullet$) with any hydrocarbon radical (R$\bullet$ ) (14) and (15)~\citep{vi,vi1,iii,vi2}; formyl itself is formed through the addition of a suprathermal hydrogen atom to carbon monoxide (13), while hydrocarbon radicals are generated through interactions of GCR proxies with closed shell hydrocarbon such as methane (1) and ethane (12). Aldehydes act as precursors to their enol tautomers, such as  vinylalcohol---also known as hydroxyethylene---(C$_2$H$_3$OH) and cis/trans 1-methyl-2-hydroxy­ethylene ((CH$_3$)HCCH(OH)), which was not identified in the present search~\citep{vi,vii,vii1}. Note that acetaldehyde (CH$_3$CHO), along with the vinylalcohol (C$_2$H$_3$OH) and propylene oxide (c-C$_2$H$_4$O), were also identified in the ethylene--carbon dioxide  ice mixture exposed to ionizing radiation; in this system, the reaction was initiated by suprathermal hydrogen atoms and additions to the carbon--carbon double bond, followed by stabilization of the complex and/or isomerization prior to complex stabilization~\citep{iii,viii}.

\begin{eqnarray}
\rm C_2H_6 \rightarrow \bullet C_2H_5 + H\bullet \\
\rm H\bullet + CO \rightarrow HCO\bullet \\
\rm \bullet CH_3 + HCO\bullet \rightarrow CH_3CHO \\
\rm \bullet C_2H_5 + HCO\bullet \rightarrow C_2H_5HCO
\end{eqnarray}
Ketones. Ketones are classified as organic compounds carrying the carbonyl (C=O) functional group, along with two hydrocarbon moieties at the carbonyl carbon atom (RCOR). In the present survey, the simplest ketone, acetone (or propanone) (CH$_3$COCH$_3$), an isomer of propanal (C$_2$H$_5$HCO), was detected. Methylketones (CH$_3$COR) can be formed via radical--radical recombination of the acetyl radical (CH$_3$CO$\bullet$)~\citep{viii,ix,ix1,ix2}–--generated via radiolysis of acetaldehyde (16)---and a hydrocarbon radical (R$\bullet$) via reaction (17)~\citep{x} Therefore, acetone (CH$_3$COCH$_3$) can be formed via the reaction of the acetyl radical plus methyl (18).
\begin{eqnarray}
\rm CH_3CHO \rightarrow CH_3CO\bullet  + H\bullet \\
\rm CH_3CO\bullet   + R\bullet \rightarrow CH_3COR \\
\rm CH_3CO\bullet   + CH_3\bullet \rightarrow CH_3COCH_3
\end{eqnarray}

Carboxylic Acids. Carboxylic acids represent organic molecules that contain the carboxyl (-COOH) functional group, with formic acid (HCOOH) and acetic acid (CH$_3$COOH), as detected in the present survey and by~\citet{2002ApJ...576..264R} being the simplest representatives of this class. In polar, water-rich ices, detailed surface science experiments have revealed that formic acid (HCOOH) can be efficiently formed in carbon monoxide (CO)--water (H$_2$O) ices through formyl (HCO$\bullet$ )--hydroxyl ($\bullet$ OH) radical--radical reactions (20) within the ices~\citep{xi}; the hydroxyl radical can be gene­rated through GCR proxy interactions with water (19). Acetic acid can be synthesized in polar water (H$_2$O)–-acetaldehyde (CH$_3$CHO) ices~\citep{ix1}. Here, acetyl (CH$_3$CO$\bullet$ ) radicals formed via radiolysis of acetaldehyde (CH$_3$CHO) (reaction (16)) undergo radical--radical recombination via reaction (21), with hydroxyl radicals (OH) being generated through radiolysis of water (19). 
\begin{eqnarray}
\rm H_2O \rightarrow \bullet OH + H\bullet \\
\rm \bullet OH + HCO\bullet \rightarrow HCOOH\\
\rm CH_3CO\bullet  + OH\bullet \rightarrow CH_3COOH
\end{eqnarray}
Acetic acid, along with more complex carboxylic acids with organic sidechains, can also be generated in apolar ices containing carbon dioxide and hydrocarbons (RH)~\citep{xiii,xiii1}. Here, a generic hydrocarbon (RH) can be radiolyzed to the organic radical plus atomic hydrogen (22), followed by the addition of suprathermal hydrogen to carbon dioxide, forming the hydroxycarbonyl (HOC$\bullet$ O) species (23). Thereafter, the hydrocarbon radical has been shown to recombine with the hydroxy­carbonyl (HOC$\bullet$ O) radical, forming the carboxylic acid (RCOOH) (24). These investiga­tions demons­trate that the polarity of the ice matrix can dramatically alter the underlying formation mechanisms of even simple COMs. 
\begin{eqnarray}
\rm RH \rightarrow R\bullet  + H\bullet \\
\rm H\bullet  + CO_2 \rightarrow HOC\bullet O \\
\rm HOC\bullet O + R\bullet \rightarrow RCOOH
\end{eqnarray}
Esters. Esters are defined as organics carrying the RCOOR’ moiety, with R and R’ resembling organic sidechains. This survey and the previous survey at 3~mm~\citep{2010ARep...54.1084K} have identified methylformic acid ester (HCOOCH$_3$) and ethylformic acid ester (HCOOC$_2$H$_5$). In principle, the formyl radical (HCO$\bullet$ ) generated via reaction (13) has been shown to react with the methoxy radical (CH$_3$O$\bullet$ ) formed through radiolysis of methanol (CH$_3$OH), yielding methylformic acid ester (HCOOCH$_3$) (reaction (25);~\citep{xiii}. It remains to be seen if the versatile route of radical--radical reactions between formyl (HCO$\bullet$ )  and an alkoxy (RO$\bullet$ ) can be expanded to form ethylformic acid ester (HCOOC$_2$H$_5$) via the reaction of formyl with ethoxy (C$_2$H$_5$O$\bullet$ ) (26). It shall be highlighted that besides the recombination of the formyl radical (HCO$\bullet$ ) with the methoxy radical (CH$_3$O$\bullet$ ) to form methylformic acid ester (HCOOCH$_3$), via reaction (25), a recombination with the hydroxymethyl radical ($\bullet$ CH$_2$OH) leads to the astronomically observed glycolaldehyde (HCOCH$_2$OH) (reaction (27));~\citep{xv,xv1,vii,xv2}.
\begin{eqnarray}
\rm HCO\bullet  + CH_3O\bullet  \rightarrow HCOOCH_3 \\
\rm HCO\bullet  + C_2H_5O\bullet  \rightarrow HCOOC_2H_5\\
\rm HCO\bullet  + \bullet CH_2OH \rightarrow HCOCH_2OH 
\end{eqnarray}
Amides. Amides, such as the observed formamide (HCONH$_2$), carry an amide (-CONHR) moiety. Detailed experiments have revealed that formamide (HCONH$_2$) can be formed through radical--radical reaction of the formyl radical (HCO$\bullet$ ) with the amino radical (NH$_2\bullet$ )~(\citep{2011ApJ...734...78J}; reaction (28)), with the latter being formed through radiolysis of ammonia (NH$_3$) (reaction (29);~\citep{xvii}). This principle can be also expanded to N-methylformamide (HCONHCH$_3$) (reaction (30);~\citep{xviii}), which was not observed in the present survey. 
\begin{eqnarray}
\rm HCO\bullet  + NH_2\bullet \rightarrow HCONH_2 \\
\rm NH_3 \rightarrow NH_2\bullet  + H\bullet \\
\rm HCO\bullet  + CH_3NH\bullet \rightarrow HCONHCH_3
\end{eqnarray}

\subsection{Formation of hydrocarbons and nitriles} \label{subsec:formation2}

 C$_3$H. In this survey, two structural isomers of the doublet tricarbonhydride (C$_3$H) radicals were observed in the gas phase: linear propynylidyne (l-C$_3$H) and cyclopropynylidyne (c-C$_3$H). Crossed molecular beam experiments coupled with electronic structure calculations revealed that both isomers can be formed easily via the barrierless and exoergic neutral--neutral reaction between ground-state carbon atoms (C($^3$P)) and acetylene (C$_2$H$_2$) in the gas phase (reaction (31);~\citep{xix,xix1,xix2,xix3,xix4,xix5}). 
\[
\begin{array}{rr}
\rm\hspace{60mm} C(^3P)+C_2H_2 \rightarrow l-C_3H\bullet  + H\bullet & \hspace{55mm} (31a)\\
\rm C(^3P) + C_2H_2 \rightarrow c-C_3H\bullet  + H\bullet &  (31b)
\end{array}
\]
C$_3$H$_2$. Two structural isomers of tricarbondihydride (C$_3$H$_2$) were observed in the present survey: the partially aromatic, cyclic cyclopropenylidene (c-C$_3$H$_2$) and the acyclic cumulene-type propadienylidene isomer (C$_3$H$_2$). Both molecules hold a C$_2v$ point group. Once again, crossed molecular beam studies merged with electronic structure calculations revealed that both isomers can be formed in the gas-phase reactions of the methylidyne radical (CH) with acetylene (C$_2$H$_2$)~\citep{xx,xx1} and of ground-state carbon atoms (C($^3$P)) plus the vinyl radical (C$_2$H$_3\bullet$)~(\citet{xxi,xxi1}; reactions (32) and (33), respectively). It shall be noted that these elementary reactions can also form both tricarbonhydride (C$_3$H) radicals through the elimination of molecular hydrogen, albeit with lower branching ratios. 
\[
\begin{array}{rr}
\rm\hspace{60mm} CH + C_2H_2 \rightarrow c-C_3H_2 + H\bullet & \qquad \hspace{55mm}(32a)\\
\rm CH + C_2H_2 \rightarrow C_3H_2 + H\bullet & \qquad(32b)\\
\rm C(^3P) + C_2H_3 \rightarrow c-C_3H_2 + H\bullet & \qquad(33a)\\
 \rm C(^3P) + C_2H_3 \rightarrow C_3H_2 + H\bullet & \qquad(33b)
\end{array}
\]
C$_4$H and C$_5$H. Crossed molecular beam studies revealed that the linear butadiynyl isomer (l-C$_4$H), which can be considered as an ethynyl-substituted ethynyl (C$_2$H$\bullet$) radical, can be formed via the barrierless and exoergic reaction of dicarbon (C$_2$) with acetylene (C$_2$H$_2$) in the gas phase (reaction (34);~\citet{xxii,xxii1}). The linear pentadiynylidyne (C$_5$H) radical was revealed to be the exclusive product of the neutral--neutral reaction between ground-state carbon atoms (C($^3$P)) and diacetylene (C$_4$H$_2$) via reaction (35);~\citep{xxiii}).
\addtocounter{equation}{3}
\begin{eqnarray}
\rm  C_2 + C_2H_2 \rightarrow C_4H\bullet + H\bullet\\
\rm C(^3P) + C_4H_2 \rightarrow C_5H\bullet + H\bullet
\end{eqnarray}
CH$_3$CCH. The methylacetylene molecule (CH$_3$CCH) represents a peculiar case, as, to date, no gas-phase reaction involving two neutral reactants has been found to form gas-phase methylacetylene. The microwave inactive allene isomer (H$_2$CCCH$_2$), however, can be formed via the neutral--neutral reaction between the methylidyne radical (CH) and ethylene (C$_2$H$_4$)~\citep{xx1}. However, methylacetylene (CH$_3$CCH) can be formed in interstellar ices through the radical--radical reaction of a methyl radical (CH$_3\bullet$) with an ethynyl radical ($\bullet$C$_2$H) (reaction (36)), with the latter being formed via interactions of GCR proxies with acetylene (C$_2$H$_2$). Once formed, the sublimation in the hot core stage can release methylacetylene into the gas phase~\citep{xxv,xxv1}.
\begin{eqnarray}
\rm CH_3\bullet + \bullet C_2H \rightarrow CH_3CCH
\end{eqnarray}
H(CC)nCN (n = 1--3).  Three cyanopolyacetylenes, organic molecules that can be derived from polyacetylene by replacing a hydrogen atom with a cyano group, have been detected in the present survey. Cyanoacetylene, cyanodiacetylene, and cyanotriacetylene can be formed easily via the gas-phase reactions of the cyano radical (CN) with acetylene (C$_2$H$_2$), diacetylene (C$_4$H$_2$), and triacetylene (C$_6$H$_2$), respectively (37)-(39)~\citep{xxvi,xxvi1,xxvi2,xxvi3}
\begin{eqnarray}
\rm CN\bullet + C_2H_2 \rightarrow H(CC)CN + H\bullet\\
\rm CN\bullet + C_4H_2 \rightarrow H(CC)_2CN + H\bullet\\
\rm CN\bullet + C_6H_2 \rightarrow H(CC)_3CN + H\bullet
\end{eqnarray}
C$_2$H$_3$CN, CH$_3$CCCN, and H$_2$CCC(CN)H. These three nitriles can be formed in analogy to the cyanopolyacetylenes through neutral--neutral reactions of the cyano radical with the corresponding hydrocarbons ethylene (C$_2$H$_4$)~\citep{xxvii}, methylacetylene (CH$_3$CCH)~\citep{xxviii,xxviii1}, and allene (H$_2$CCCH$_2$)~\citep{xxviii1}, as verified via crossed molecular beam studies (reactions (40)-(42)). 

\begin{eqnarray}
\rm CN\bullet + C_2H_4 \rightarrow C_2H_3CN + H\bullet\\
\rm CN\bullet + CH_3CCH \rightarrow CH_3CCCN+ H\bullet\\
\rm CN\bullet + H_2CCCH_2 \rightarrow H_2CCC(CN)H + H\bullet
\end{eqnarray}

CH$_3$CN and C$_2$H$_5$CN: In analogy to methylacetylene, no gas-phase synthesis has been verified so far leading to the formation of methylcyanide (CH$_3$CN) and ethylcyanide (C$_2$H$_5$CN) involving neutral--neutral reactions. However, these saturated nitriles can be formed within interstellar grains via radical--radical recombination of the cyano radical with the methyl and ethyl radicals, respectively, via reactions (43) and (44); it is important to note that this route has not been verified unequivocally in laboratory experiments to date.  

\begin{eqnarray}
\rm CN\bullet + CH_3\bullet \rightarrow CH_3CN \\
\rm CN\bullet + C_2H_5\bullet \rightarrow C_2H_5CN
\end{eqnarray}

\section{Conclusion}

A spectral-line survey of the massive star formation region W51e1/e2 at 68--88 GHz was conducted with the 20~m radio telescope of the Onsala Space Observatory. An rms sensitivity of 0.003~K was achieved for 75\% of the observed band. A number of lines that are absent from the Lovas list of molecular lines observed in space were detected, and most of these were identified. 79~molecules and their isotopic species were detected, from simple diatomic or triatomic molecules, such as SO, SiO, and CCH, to complex organic compounds, such as CH$_3$OCH$_3$ or CH$_3$COCH$_3$. The list of the detected molecules consists of both the parent species and hydrogen-, carbon-, oxygen-, and nitrogen-related isotopologues. In addition to neutral molecules, various molecular ions were detected. Some of these (HC$^{18}$O$^+$, H$^{13}$CO$^+$, and HCS$^+$) usually exist in molecular clouds with high visual extinctions $A_V$. 

A significant number of the detected molecules are typical for hot cores. These include the neutral molecules HCOOCH$_3$, CH$_3$CH$_2$OH, and CH$_3$COCH$_3$, which are currently believed to exist in the gas phase only in hot cores and shock-heated gas. Apart from the rotational lines in the ground vibrational state, vibrationally excited lines of several molecules were detected. Among them were C$_4$H and HC$_3$N lines, with upper-level temperatures of several hundred Kelvins. Such lines can arise only in hot gas, with temperatures of the order of 100 K or higher. 10 of 15 plausible RDs yield rotational temperatures on the order of, or greater than, 100~K. Therefore, it is obvious that the emissions of many detected molecules arise just in the hot cores e2 and e8. Unfortunately, we could not separate the contributions of these objects.

Nevertheless, other molecules (CH$_3$CCH, HC$_5$N, CH$_2$CO, etc.) demonstrate lower rotation temperatures, showing that the bulk of their emission arises in colder gas in the vicinity of hot cores.

Potential formation pathways of COMs and of hydrocarbons, along with nitriles, are considered. These formation routes are first discussed in the context of laboratory experiments elucidating the synthesis of organic molecules in interstellar ices in cold molecular clouds, followed by sublimation into the gas phase in the hot core stage. Thereafter, we discuss the predominant formation of hydrocarbons and their nitriles in the gas phase through bimolecular neutral--neutral reactions.

\begin{acknowledgments}
The studies at Lebedev Physical Institute were supported by the Ministry of Science and Higher Education of the Russian Federation by the grant No. 075-15-2021-597. 
This paper makes use of NASA's Astrophysical Data System as well as of the following data set: Observed Interstellar Molecular Microwave Transitions~\citep{lovas09}. 
The Onsala Space Observatory national research infrastructure is funded through Swedish Research Council grant No. 2017-00648.
\end{acknowledgments}

\facility{OSO: 20m (Chalmers University of Technology 20m Telescope at Onsala Space Observatory)}
\software{GILDAS: Grenoble Image and Line Data Analysis Software\\
(Gildas Team, https://www.iram.fr/IRAMFR/GILDAS)}

\end{document}